\documentclass[twocolumn]{IEEEtran}

\usepackage[utf8]{inputenc}
\usepackage[ruled,vlined]{algorithm2e}
\usepackage{amsmath}
\usepackage{amsthm}
\usepackage{amsfonts}
\usepackage{mathrsfs}
\usepackage{pifont}
\usepackage{amssymb}
\usepackage{verbatim}
\usepackage{upgreek}
\usepackage{color}
\usepackage{epsfig}
\usepackage{booktabs}
\usepackage{bm}
\usepackage{setspace}
\usepackage[hyphens]{url}
\usepackage{graphicx}
\usepackage[overload]{empheq}
\usepackage{fancyhdr}
\usepackage[bookmarks=true,hidelinks,pdfpagelabels=true]{hyperref}
\usepackage{cite}
\usepackage{float}

\usepackage[crop=pdfcrop,cleanup={.tex, .dvi, .ps, .pdf, .log, .bbl, .out, .upa, .upb}]{pstool}
\usepackage[caption=false,font=footnotesize]{subfig}
\mathtoolsset{showonlyrefs = true}
\usepackage{tikz}
\usetikzlibrary{shapes,arrows,calc}
\usetikzlibrary{positioning}

\usepackage{balance}

\newcommand{\rd}{}

\newcommand{\T}{\text{T}}

\newcommand{\norm}[1]{\left\lVert#1\right\rVert}

\newcommand{\V}[1]{\bm{#1}}

\newcommand{\Set}[1]{{\mathcal{#1}}}

\newcommand{\br}{\mbox{\boldmath{$r$}}}

\newcommand{\bt}{\mbox{\boldmath{$t$}}}

\newcommand{\bx}{\mbox{\boldmath{$x$}}}

\definecolor{myBlue}{rgb}{0,0.45,0.74}
\definecolor{myGreen}{rgb}{0.47,0.67,0.19}
\tikzstyle{int} = [draw, fill = gray!20, minimum height = 1.0cm, minimum width = 5cm, align = center, inner sep = 2mm, rounded corners = 1ex, font=\small]
\tikzstyle{intBlue} = [draw = myBlue, fill = white, very thick, minimum height = 1.0cm, minimum width = 4cm, align = center, inner sep = 2mm, rounded corners = 1ex, font=\small]
\tikzstyle{intGreen} = [draw = myGreen, fill = white, very thick, minimum height = 1.0cm, minimum width = 4cm, align = center, inner sep = 2mm, rounded corners = 1ex, font=\small]
\tikzstyle{int2} = [draw, fill = gray!20, minimum height = 1.0cm, minimum width = 3.5cm, align = center, inner sep = 2mm, rounded corners = 1ex, font=\small]

\definecolor{OliveGreen}{rgb}{0,0.6,0}

\newcommand{\paperTitle}{3D Localization and Tracking Methods for Multi-Platform Radar Networks\vspace{1.2mm}}

\allowdisplaybreaks

\newcommand{\deq}{\triangleq}

\begin{document}

\title{\paperTitle}

\author{Angela~Marino, Giovanni~Soldi, Domenico~Gaglione, Augusto~Aubry, Paolo~Braca,\\
Antonio De Maio, and Peter Willett

\thanks{A. Marino
is with the Department of Electrical Engineering and Information Technology (DIETI) of the University of Naples Federico II, Italy (e-mail: angela.marino@unina.it).}
\thanks{A. Aubry and A. De Maio are with the Department of Electrical Engineering and Information Technology (DIETI) of the University of Naples Federico II, Italy, and with the National Inter-University Consortium for Telecommunications, Parma, Italy (e-mail: \{augusto.aubry, antonio.demaio\}@unina.it).}
\thanks{G. Soldi, D. Gaglione, P. Braca are with the NATO STO Centre for Maritime Research and Experimentation (CMRE), La~Spezia, Italy (e-mail: \{giovanni.soldi, domenico.gaglione, paolo.braca\}@cmre.nato.int).}
\thanks{P.\ Willett is with the University of Connecticut, Storrs, CT 06269, USA (e-mail: peter.willett@uconn.edu).}
\thanks{This work was supported in part by the NATO Allied Command Transformation (ACT) under project DKOE.}
}
		
\maketitle

\begin{abstract}
Multi-platform radar networks (MPRNs) are
an emerging sensing technology due to their ability to provide improved surveillance capabilities over plain monostatic and bistatic systems.
The design of advanced detection, localization, and tracking algorithms for efficient fusion of information obtained through multiple receivers has attracted much attention. However, considerable challenges remain.
This article provides an overview on recent unconstrained and constrained localization techniques as well as multitarget
tracking (MTT) algorithms tailored to MPRNs.
In particular, two data-processing methods are illustrated  and explored in detail, one aimed at accomplishing localization tasks the other tracking functions.
As to the former, assuming a MPRN with one transmitter and multiple receivers, the angular and range constrained estimator (ARCE) algorithm
capitalizes on the knowledge of the transmitter
antenna beamwidth. As to the latter, the scalable sum-product algorithm (SPA) based MTT technique is presented.
Additionally,
a solution to combine ARCE and SPA-based MTT is investigated in order to boost
the accuracy of the overall surveillance system. Simulated experiments show the
benefit of the combined algorithm in comparison with the conventional baseline SPA-based MTT and the stand-alone ARCE localization, in
a 3D sensing scenario.
\end{abstract}

\begin{IEEEkeywords}
One-shot target localization, angular and range constrained estimator, multi-target tracking, particle filtering, sum-product algorithm. 
\end{IEEEkeywords}

\section{Introduction}

\subsection{Multi-platform radar networks: An overview}
\label{sec:multi_plat_overview}

{Modern surveillance systems encompass the use of multiple cooperative, autonomous, and unmanned vehicles (see, e.g.,~\cite{BraGagSolMenFerLepNicWilBraWin,AUV_Access_2019,hassanalian2017classifications,ferri2017cooperative,Zeng_16,DIFFUSION2015}) in all the different operational domains, that is, air, land and sea~\cite{de2021understanding,Pastore_2017}.
In this context} a multi-platform radar network (MPRN) is envisioned as a next-generation sensing system capitalizing on several spatially separated transmitting, receiving and/or transmitting-receiving deployable sensing units.
{Data acquired and pre-processed} by the nodes can be either suitably shared among them or, in a joint configuration, sent to a central fusion node, which fully processes {the collected information} implementing detection, localization, and tracking algorithms \cite{BraGagSolMenFerLepNicWilBraWin}.
MPRNs have gained increasing interest in the last decades as they can provide relevant performance improvements over monostatic and bistatic (i.e., with one transmitter and one receiver) radars \cite{Che:B98,OgaDouIng:B18}.
The use of several nodes allows to widen the coverage area and to \textit{look} at the targets 
from different aspect angles, thus enhancing the targets monitorability via spatial diversity. 
Moreover a great advantage of such systems
is cognitive superiority~\cite{hartley2020cognitive, AubryCDP19_55}, achieved by means of their collaborative self-coordination aimed at a fully-optimized and appropriate environment adaptation~\cite{WicMoo}.

\begin{figure}[!t]
    \centering
    \includegraphics[width=1\columnwidth]{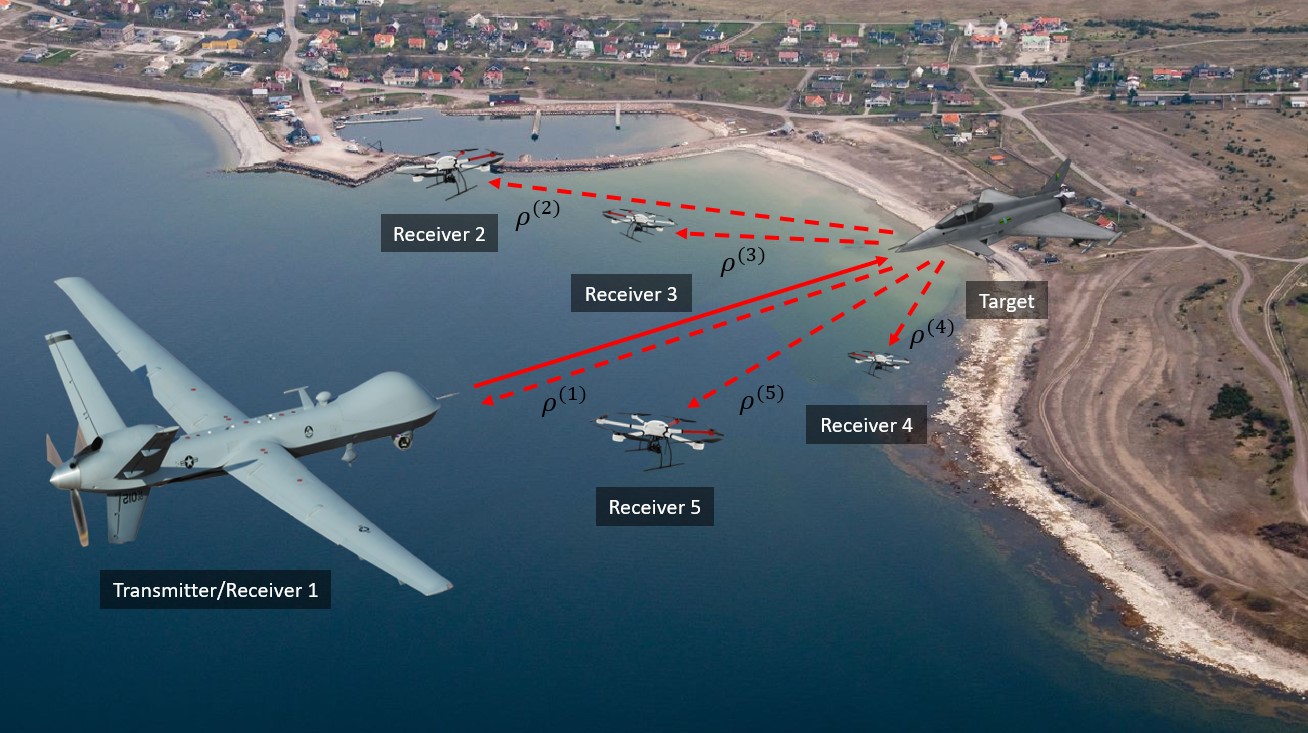}
        \caption{Notional representation of a
                MPRN with one transmitting-receiving node and four receiving nodes.
                }
    \label{fig:multiplatform_system}
\end{figure}
Such cooperation demands a tight synchronization among the nodes which can be achieved through the use of global positioning system (GPS) and highly stable GPS disciplined oscillators (GPSDOs)\cite{Lomb,SanIng}.
Further flexibility and performance improvements can be achieved via geometric diversity, namely  dynamically optimizing the number and the locations of the individual platforms \cite{GriBak, AubryBDFS23_71} as, for instance, unmanned aerial vehicles (UAVs) deployable on the base of a specific task.
Moreover, MPRNs are capable of reducing shadowing effects, which are particularly damaging in urban
environments,
and they are naturally more robust to jamming thanks to their topology.
These advantages come at the expense of an overall increased complexity in designing and managing these systems, e.g., necessity of data transmission links and of synchronization among the sites, choice of operational frequencies and waveforms for different transmitters and receivers, definition and implementation of bespoke cooperation and scheduling strategies, just to mention a few~\cite{DeMaioBook, YiangADYC22_70}.

Prototypes of ground-based
MPRNs are the NetRAD \cite{DerDouWooBak:J07}
and its evolution NeXtRAD \cite{IngGriFioRitWoo:C14,IngLLewPalRitGri:C19}.
NetRAD,
designed by the University College London and the University of Cape Town,
is a low-cost, coherent, short-range (up to 1 km) pulse Doppler MPRN 
composed
of three
transmitting-receiving nodes operating over a 50 MHz bandwidth
in S-band, and wire connected
for time and
phase control, as well as synchronization and data sharing.
NetRAD was proved effective in detecting and localizing multiple moving targets, with a valuable performance boost as compared to monostatic and bistatic systems \cite{DouWooBak:C07}, leveraging clutter diversity~\cite{Griffiths14, NickelGLGK17_2, AubryCDF23_59}.
NeXtRAD is a three nodes (one transmitting-receiving sensor, also acting as fusion centre, and two receiving units), dual band (X and L), fully polarimetric MPRN; the active node can
transmit either horizontally or vertically polarized pulses in either X- or L-band and the receiving nodes can possibly 
acquire signals simultaneously on both polarizations \cite{BeaRitGriMicIngLewKah:C20}.
The nodes are connected through WiFi, thus permitting a wider spatial separation and a faster deployment of the overall system.
NeXtRAD,
along with its predecessor NetRAD,
has allowed the collection of a significant database of multistatic and multipolarimetric measurements \cite{BeaRitGriMicIngLewKah:C20}.
In the maritime domain, a ground-based and deployable MPRN has been developed and tested at the NATO Centre for Maritime Research and Experimentation (CMRE). This MPRN is a coherent high-resolution X-band radar network consisting of two radar nodes operating simultaneously in multistatic configuration: a maritime radar and an inverse synthetic aperture radar (SAR) node \cite{VivBraGraWil}. In this context novel random matrix models have been proposed to develop extended target tracking approaches, see, e.g., \cite{VivBraGraWil2} and references therein. Furthermore, a novel extended target detection methodology based on machine learning has been proposed in~\cite{BracaMAMMW22}.   

\vspace{2mm}

\begin{figure*}[!t]
	\centering
	\begin{tikzpicture}
		\node[] at (0,0) {\includegraphics{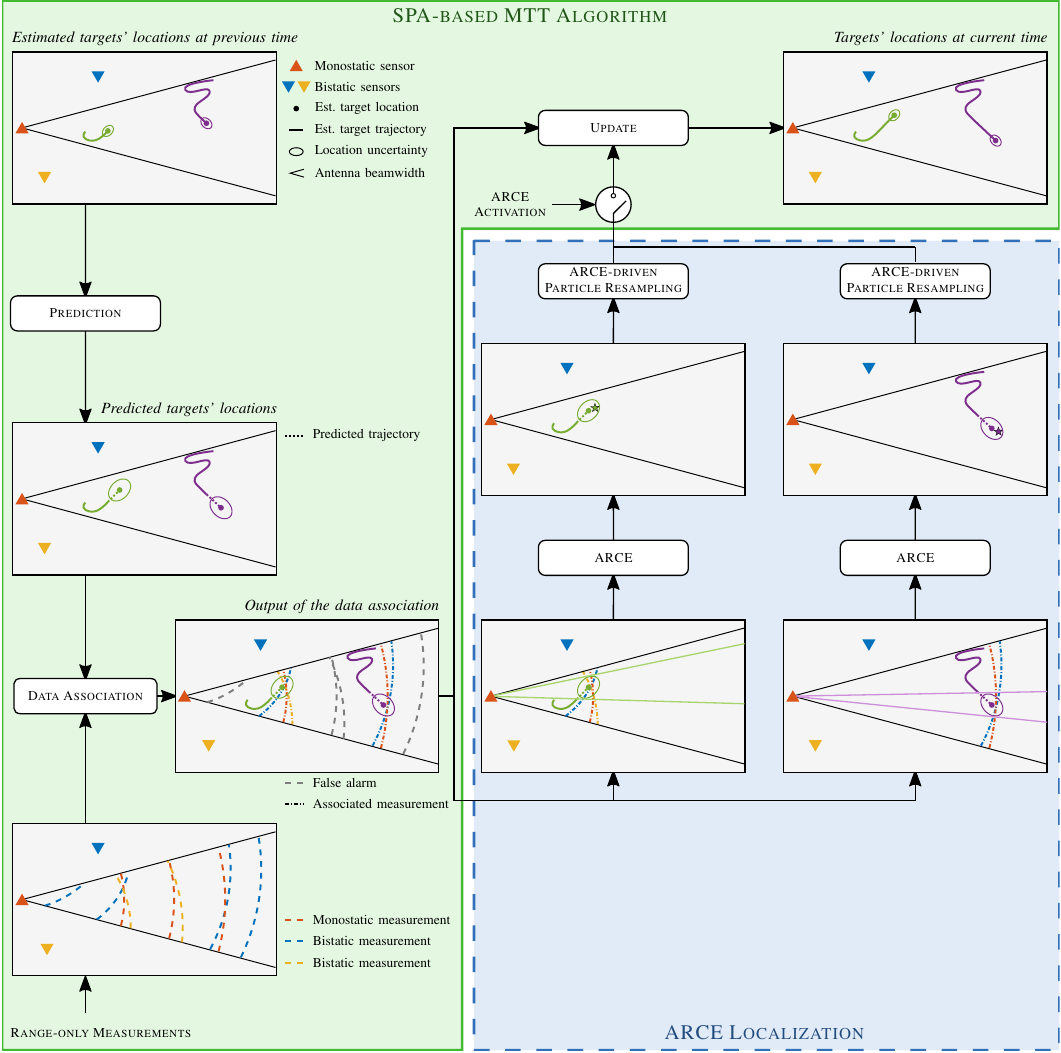}};
		
				\draw[semithick, ->] (1.35,1.99) -- (1.9,1.90);
		\draw (2.45,1.85) node {			\begin{minipage}{0.05\textwidth}
				\tiny\centering ARCE Estimate
			\end{minipage}
		};
		
				\draw[semithick, ->] (8.05,1.75) -- (8.22,2.05);
		\draw (8.25,2.36) node {			\begin{minipage}{0.05\textwidth}
				\tiny\centering ARCE Estimate
			\end{minipage}
		};

				\node[] at (4,-6.84){			\begin{minipage}{0.53\textwidth}
				\footnotesize
				\begin{spacing}{1.15}
					Localization is performed through ARCE. Exploiting the prior information about the antenna beamwidth of the transmitter, i.e., the black solid lines, and the \textit{virtual} beamwidth obtained from the target predicted uncertainty, i.e., the colored (green or purple) solid lines, ARCE provides a location estimate of each target, depicted as stars in the top images.
					The ARCE estimate of each target is then used to extract an additional set of particles that \textit{replaces} a subset of predicted particles; details are provided in Section~\ref{sec:ARCE_plus_SPA}. The newly ARCE-driven sets of particles are eventually used --- if the switch is \textit{on} --- for the update of the targets' locations at current time.
                        				\end{spacing}
			\end{minipage}
		};
	\end{tikzpicture}
	
	\caption{
         Green solid
                  area: notional sketch of the steps of the SPA-based MTT algorithm for a MPRN in a 2D scenario, representing only target locations, instead of the full target states. Joint green solid and blue dashed
                  areas: proposed localization-enhanced MTT algorithm for a MPRN.}
	\label{fig:notional-sketch}
 
\end{figure*}
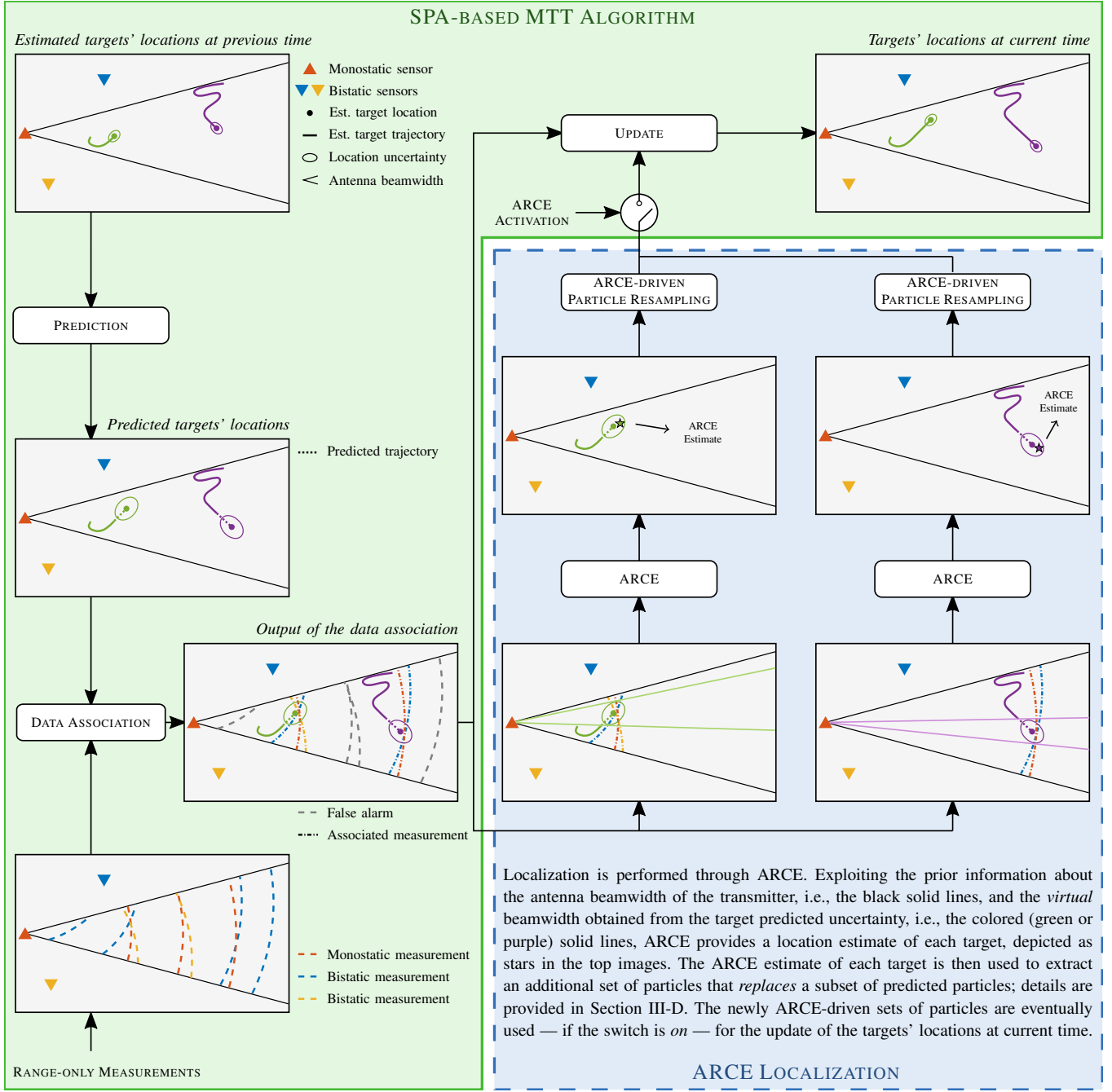

\subsection{Brief Description of the Main Framework and Paper Contribution}
\label{sec:cont_paper_org}
The main aim of this article is to revisit and
tailor multitarget tracking (MTT) algorithms for MPRNs with a single transmitting node (extensions to multiple transmitting 
devices is straightforward) and multiple receiving units {(one of them co-located with the transmit node, i.e., operating as a monostatic radar)} that collect time-of-arrival (ToA) measurements or, equivalently, bistatic range measurements,
in a 3D scenario. These 
algorithms aim
{at} sequentially
{determining} the unknown number of targets in a surveilled area as well as to estimate their states, e.g., positions and velocities, 
using the bistatic measurements collected
by multiple receivers up to the current time.
Here,
emphasis is given to
a particle-based Bayesian MTT method
that exploits the sum-product algorithm (SPA)
\cite{MeyBraWilHla:J17,MeyKroWilLauHlaBraWin:J18};
the main steps of this algorithm for a MPRN in a simple 2D scenario are
{enclosed} in the green solid area of Fig.~\ref{fig:notional-sketch}.
{To grasp further insights on this processing framework, let} us consider a MPRN
{including} a single
{monostatic sensor (represented by a red triangle) with} a limited antenna beamwidth,\footnote{Note that, we generally refer as the antenna or radar beamwidth to the antenna 3dB beamwidth.} and
{two} {passive (receiving only)} 
{ nodes}
located elsewhere {(represented by a blue and a yellow triangles).}
Moreover, let us suppose that two targets are moving within the monostatic sensor's antenna beamwidth and are currently tracked.
Targets' locations --- green and 
purple dots ---
at current time
{are related to those at previous time through} a kinematic model that properly describes the dynamic of the targets.
Each sensor produces range-only measurements of the targets, if these are detected, as well as unwanted measurements due to clutter, all represented by dashed lines.
Then, {through a procedure called data association,} the measurements are either associated to the
supposedly extant targets, or treated as false alarms (i.e., clutter-generated), or assumed to be produced by newly observed targets.
The associated measurements are eventually used to update the targets' locations at current time.

Compared to other state-of-the-art 
MTT methods, the SPA-based MTT algorithm is characterized by exceptional scalability in terms of
computational complexity with respect to the number of receivers, targets, and 
measurements.
{Nevertheless, in a 3D scenario a large number of particles might be required to efficiently sample the target space, thus increasing the computational burden. This is particularly evident for the initialization of a new target track from range-only measurements. Indeed, the lack of angle information requires the prior distribution of the target state --- comprising both 3D position and 3D velocity --- to cover a large volume, which cannot be reliably represented by a small number of particles. As consequence, the resulting imprecise representation of the prior target state distribution can propagate via the target dynamic  over time, eventually making the overall target tracking inaccurate.}

In order to mitigate the aforementioned shortcomings, in this paper we
{propose to boost} the performance of the SPA-based MTT algorithm
capitalizing on 
{single-snapshot} localization algorithms. These algorithms are able to estimate the target location by {just} using the associated measurements (i.e., after the data association) already available at the MTT algorithm.
Different unconstrained and constrained localization
methods are available in literature, and
here we
focus on the angular and range constrained estimator (ARCE) recently proposed in \cite{AubBraDemMar:J22}, which leverages the prior information about the antenna beamwidth of the transmitter.
Precisely, the main contribution of this paper is the 
embedding within the SPA-based MTT method of the ARCE estimate to enable a more effective sampling of the target state space, in particular during the initialization phase and when
a low number of particles is used.

Indeed the ARCE location 
estimate, being unaffected by the kinematic model and depending 
only on the measurements, can suggest a better sampling of the 
target state space and thus mitigate the negative effects of a 
rough representation of the target distribution due to particle sparsity.
The synergy between the SPA-based MTT algorithm and the
single-snapshot localization
through ARCE is sketched in
Fig.~\ref{fig:notional-sketch}, looking at both the green solid area and the
blue dashed area.
ARCE takes as input the predicted location of each
target and its associated measurements, and
exploiting the available information about both the antenna 
beamwidth of the transmitter (i.e., the black solid lines) and a 
bespoke  \textit{virtual} beamwidth 
 (i.e., the colored, green or purple, solid lines), 
provides a location estimate of each target, depicted as stars.
Specifically, the virtual beamwidth is obtained as the intersection of the physical antenna beamwidth of the transmitter and a tailored beam. This tailored beam is steered towards the target predicted position and its width is proportional to the target predicted uncertainty.
Hence, an ARCE-driven particle resampling strategy  
is aimed at replacing a subset of predicted particles with a new set of particles drawn from a distribution whose parameters depend on the performed ARCE location estimate;
the newly ARCE-driven set of particles is eventually used for the update of the target's location at current time, provided that the switch is turned on by the ARCE activation flag.\footnote{The ARCE activation flag is a  pre-fixed parameter either set to true, if ARCE is used by the SPA-based MTT algorithm or false, otherwise.}

The remainder of the paper is organized as follows.
Section~\ref{sec:Single_target_localization} provides a survey on uncostrained and constrained
{single-snapshot} single target localization algorithms; among them, ARCE is explored.
MTT algorithms for MPRNs and, in particular, the SPA-based MTT algorithm and the SPA-based MTT algorithm enhanced by ARCE are described in Section~\ref{sec:MTT_algorithms}.
Finally, simulation results and conclusions are reported in Section~\ref{sec:simulated_experiments} and Section~\ref{sec:conclusions}, respectively.

\section{{Single-Snapshot} Single-Target Localization}
\label{sec:Single_target_localization}

\begin{figure*}[!t]

	\centering
	\subfloat[Noise-free scenario. Time scan $k = 1$.]{
		\includegraphics{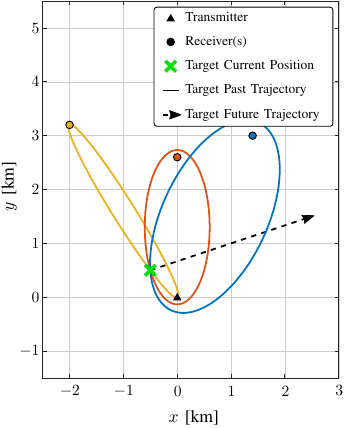}
		\label{fig:ellipse-noise-free-step-1} 
	}	
	\centering
	\subfloat[Noise-free scenario. Time scan $k = 2$.]{
		\includegraphics{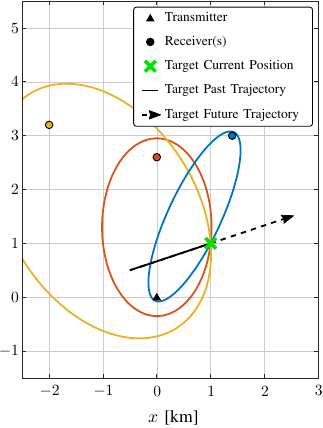}
		\label{fig:ellipse-noise-free-step-2} 
	}	
	\centering
	\subfloat[Noise-free scenario. Time scan $k = 3$.]{
		\includegraphics{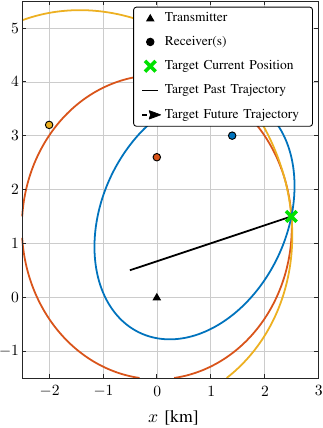}
		\label{fig:ellipse-noise-free-step-3} 
	}
	
		\centering
	\subfloat[Noisy scenario. Time scan $k = 1$.]{
		\includegraphics{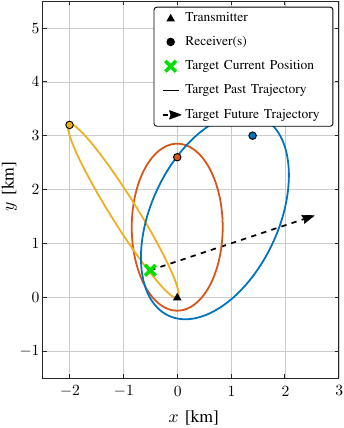}
		\label{fig:ellipse-noisy-step-1} 
	}	
	\centering
	\subfloat[Noisy scenario. Time scan $k = 2$.]{
		\includegraphics{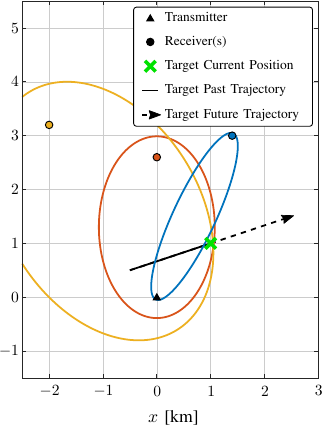}
		\label{fig:ellipse-noisy-step-2} 
	}	
	\centering
	\subfloat[Noisy scenario. Time scan $k = 3$.]{
		\includegraphics{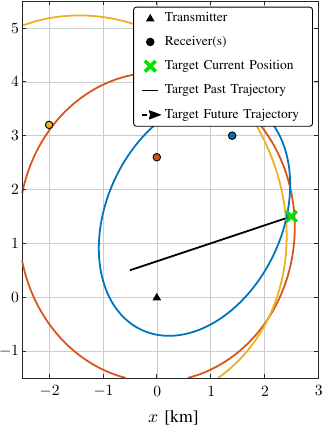}
		\label{fig:ellipse-noisy-step-3} 
	}
      
	\caption{Notional representation of the target localization process in correspondence of three different time-instants, assuming three stationary transmitter-receiver baselines.
	Panels (a), (b), and (c) refer to a noise-free scenario and the target position is identified by the unique intersection among the three ellipses. Panels (d), (e), and (f) refer to noisy measurements and there is not a unique intersection among the three ellipses.} 
	\vspace{-2mm}
	\label{fig:ellipse}
\end{figure*}

Let us consider a MPRN comprising one transmitter and $S$ receivers, whose task is the localization of a prospective moving target present in its coverage area.
In the ideal conditions of perfect target detectability, no false alarms, and noise-free measurements,
the target position can be exactly
determined exploiting an adequate number of receivers  \cite{WilGri:B07,MalKul:J12,OgaDouIng:B18}.
To shed light on the localization process, let us focus on the estimation of the target
position in a 2D scenario at the $k$-th snapshot and indicate  with $\bt_k$ and $\br_k^{(i)}$
the positions of the transmitter and the $i$-th receiver, respectively, while
$\bx_k$ denotes the \textit{unknown} target position. 

Before proceeding further, it is worth mentioning that {single-snapshot} target localization just relies on data (at all the receiving nodes) from a single time-instant/snapshot. As a consequence, it neither takes advantage of previously collected data/information, nor
does it exploit any knowledge
of the target dynamics.\footnote{Note that, some localization methods relying on previous data and dynamics modelling have been also proposed in open literature \cite{GiaCecScoGar,UllSheSuEspCho}.}
In such a context, the noise-free bistatic range measurement (assuming an ideal synchronization among the nodes) at time instant $k$ collected by the  $i$-th receiver node --- as triggered by its detection process --- is provided by the following expression
\begin{align}
	\rho_k^{(i)} =  \| \bx_k - \bt_k \| + \| \bx_k - \br_k^{(i)} \| \, , \quad i = 1, \ldots, S \, .
	\label{eq:bistatic-range-noise-free}
\end{align}
The $i$-th bistatic range measurement in (\ref{eq:bistatic-range-noise-free}), along with transmitter and receiver positions, identifies an ellipse (ellipsoid in a 3D geometry) whose foci are
located at $\bt_k$ and $\br_k^{(i)}$, and whose
major axis\footnote{When the transmitter is co-located with the $i$-th receiver then the $i$-th ellipse boils down to a circle centered at $\bt_k=\br_k^{(i)}$. The resulting Tx-Rx pair corresponds to a monostatic radar.} 
is ${\rho_k^{(i)}}$.
Since the target position has to simultaneously fulfill the $S$ relationships described by \eqref{eq:bistatic-range-noise-free}, $\bx_k$ can be determined, in general, as the intersection of $S$ ellipses;
specifically, in the 2D scenario, three receivers are sufficient to accomplish the task as long as the transmitter-receiver pairs are not collinear.
Figs.~\ref{fig:ellipse}(a), \ref{fig:ellipse}(b), and \ref{fig:ellipse}(c) illustrate the target localization process (for a noise-free scenario) in correspondence of three time
scans, respectively, where the underlying target trajectory is depicted as a dashed line.  Therein, one stationary transmitter is located at the origin (i.e., $\bt_k = \bt = [0 \,\, 0]^{\T}$), and $S = 3$ stationary receivers are placed at $\br^{(1)} = [0 \,\, 2.6]^{\T}$ km, $\br^{(2)} = [-2 \,\, 3.2]^{\T}$ km, and $\br^{(3)} = [1.4 \,\, 3]^{\T}$ km.

In practice, the gathered bistatic ranges are affected by measurements noise and the resulting acquired data (embedding location information) can be modeled as
\begin{align}
	\hspace{-2mm}\rho_k^{(i)} =  \| \bx_k - \bt_k \| + \| \bx_k - \br_k^{(i)} \| + w^{(i)}_k \, , \quad i = 1, \ldots, S \, ,
	\label{eq:bistatic-range-noisy}
\end{align}
where $w^{(i)}_k$, $i=1,\cdots,S$, are zero-mean
(usually Gaussian distributed) random variables independent across $i$ and $k$.
Figs.~\ref{fig:ellipse}(d), \ref{fig:ellipse}(e), and \ref{fig:ellipse}(f) refer to the same localization scenarios of Figs.~\ref{fig:ellipse}(a), \ref{fig:ellipse}(b), and \ref{fig:ellipse}(c), respectively, but for the presence of noisy bistatic range measurements.
Inspection of the figures clearly show that the resulting ellipses
no longer intersect
at a single point, implying that a
simple process of ellipse intersection is inadequate. 

In this respect, in the following subsections, an overview on unconstrained and constrained localization techniques based on noisy bistatic range measurements is provided.

\subsection{An Overview on Unconstrained Localization Methods}
\label{subsec:unconst_sing_tar_localization}

Several methods have been proposed in the open literature to handle the (bistatic) range-only localization task in the presence of measurement errors, see e.g.~\cite{BisFidAndDogPat}.
For instance, the maximum likelihood (ML) method can be employed to locate the target iteratively with
an initial position estimate. Alternatively, the set of nonlinear
equations \eqref{eq:bistatic-range-noisy} can be converted into a set of linear ones;
thus, the least squares (LS) framework can be exploited (along with some structural constraints) to estimate (in a sub-optimal way) the target location from the linearized equations. 

From a practical point of view, it is very important to account for the
trade-off between localization accuracy and computational cost, for the selection of the appropriate method.
Indeed, while ML
provides consistent and asymptotically efficient estimators, LS, on the other hand, provides computationally affordable estimates.

In \cite{OguGomXvStoOli}, the problem of locating a single source from range measurements to a set of nodes in a wireless sensor network is addressed. 
Two localization techniques for Gaussian or Laplacian noise, respectively,  are
designed according to the ML criterion, which are based on the convex relaxation of the corresponding likelihood functions.
The proposed algorithms
exhibit appealing tradeoffs between localization accuracy and computational cost.

An LS approach to 2D target localization
yielding a closed-form solution is proposed in \cite{DiaTabDiaSed}.
Therein linearization is obtained by selecting the
first transmitter and the first receiver as primary reference, defining some proper auxiliary variables, and through algebraic computations.
Finally, by applying the LS estimation to the obtained system of linear
equations
the location estimate is
computed.
In \cite{AmiBehZam}, the problem of 2D/3D target localization from
bistatic range measurements in distributed multiple-input multiple-output (MIMO) radar systems is investigated.
The variance of the measurements
is assumed to be dependent
on the corresponding transmitter-to-target and target-to-receiver distances. By introducing auxiliary parameters, a pseudolinear set of range equations is established.
Then, the positioning problem is  solved in closed-form by a
multistage weighted LS (WLS) estimator.

\subsection{Brief Description of the Angular and Range Constrained Estimator}
\label{subsec:const_sing_tar_localization}

A viable means to improve localization performance is to capitalize on suitable a-priori information related to the features of the MPRN, such as the actual illuminated areas.
This is exactly the rationale behind the development of the method in \cite{AubBraDemMar:J22}, referred to as ARCE, where a-priori target angular information\footnote{Following similar design guidelines, advanced localization algorithms for 2D passive bistatic radars (PBRs) have been devised in \cite{AubCarDemPal:J20,AubBraDemMar:J21}.}
is exploited to improve target localization in a MPRN with one transmitter and multiple receivers.\footnote{One of the receivers, co-located with the transmitter, actually 
establishes a monostatic radar.}
Specifically, ARCE
leverages the available information about the features of the monostatic radar radiation pattern as well as its collected measurements
to enforce bespoke angular and range constraints in the target positioning process.

Without loss of generality, the receiver, labeled with $i = 1$, is co-located with the transmitter, i.e., $\V{t}_k = \V{r}^{(1)}_k$ for all $k$, and thus represents the monostatic active radar.

The ARCE algorithm estimation process is devised as the solution of a constrained LS problem, where the constraints 
force the target
to lie within the area illuminated by the monostatic active radar, and the objective function is a squared norm cost function taking into account noise acquisitions and measurement model.
Resorting to the Karush-Kuhn-Tucker (KKT) conditions, the optimal solution is obtained in a quasi-closed form\footnote{The evaluation of the candidate optimal solutions involves only elementary functions and roots of polynomial equations. The overall computational complexity of ARCE is proportional to the squared number of receivers.} following the procedure 
sketched in Fig.~\ref{fig:block-diagram} and described below (details are given in \cite[Prop. III-1]{AubBraDemMar:J22}):
\begin{enumerate}
    \item \textit{Partitioning of the  feasible target positions set}:
     leveraging the angular and range constraints, 
    the set of feasible target locations is partitioned into six
    subsets;
        \item \textit{Evaluation of candidate optimal solutions}:
        exploiting the KKT optimality conditions \cite{Ber:B16}, a finite number of (almost in closed-form) candidate optimal solutions is determined for each subsets. To this end, a smart rooting process, outlined in \cite[Sec. III-A]{AubBraDemMar:J22}, can be exploited.
     	\item \textit{Selection of the optimal solution}: among the optimal candidates--- at most twenty-six ---  the point achieving the lowest value of the objective function is selected as the target position estimate. 
\end{enumerate}
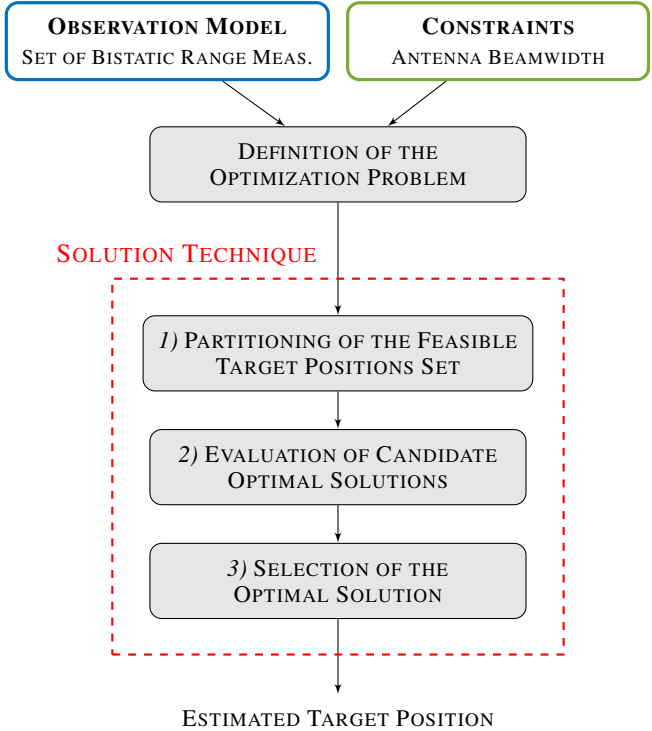
\begin{figure}[!h]
    \centering%
    \begin{tikzpicture}[auto,>=latex']

        \node [int] (a) [] {\textsc{Definition of the} \\ \textsc{Optimization Problem}};
        
        \node [intBlue] (obs) [above left=0.6cm and -2.4cm of a] {\textbf{\textsc{Observation Model}} \\[0.5mm] \footnotesize{\textsc{Set of Bistatic Range Meas.}}};
        
        \node [intGreen] (con) [above right=0.6cm and -2.4cm of a] {\textbf{\textsc{Constraints}} \\[0.5mm] \footnotesize{\textsc{Antenna Beamwidth}}};
        
		\path[->] (obs) edge node {} (a);
		\path[->] (con) edge node {} (a);
        
        \node [int] (b) [below=1.5cm of a] {\textsc{\textit{1)} Partitioning \rd{of the} Feasible} \\ \textsc{Target Positions Set}};
        \node [int] (c) [below=.5cm of b] {\textsc{\textit{2)} Evaluation of Candidate} \\ \textsc{Optimal Solutions}};
        \node [int] (d) [below=.5cm of c] {\textsc{\textit{3)} Selection of the} \\ \textsc{Optimal Solution}};
        \node [inner sep=2mm, font=\small] (e) [below=1cm of d] {\textsc{Estimated Target Position}};
		\path[->] (a) edge node {} (b);
		\path[->] (b) edge node {} (c);
		\path[->] (c) edge node {} (d);
		\path[->] (d) edge node {} (e);

		\node[rectangle,draw] (r) [red, dashed, thick, below = 1.0cm of a, minimum height = 5.0cm, minimum width = 6.0cm] {};	
		\node[] at ($(r.north west) + (1.0cm,.3cm)$) {\textcolor{red}{\textsc{Solution Technique}}};
		
    \end{tikzpicture}
    \caption{Block diagram describing ARCE.}
    \label{fig:block-diagram}
\end{figure}%
Notably, in \cite{AubBraDemMar:J22} it is shown that ARCE is capable of outperforming modified versions of the two-step estimation (TSE) method~\cite{SheMolSal:J12} and  \cite{ZhaHo:J19}, as well as other valuable unconstrained techniques. This last observation justifies the major attention paid
to ARCE in this paper.

Before concluding this section, it is worth stressing that one-shot target localization methods rely on single time 
scan measurements without taking advantage of past information; moreover the localization procedure handles a single target in conditions of perfect detectability, i.e., neither missed detections nor clutter-generated measurements are supposed.
In the next section, MTT algorithms for MPRNs are considered, which capitalize past information via a sequential estimation process, often referred to as filtering. Multiple targets over time, i.e., across multiple time steps, are also accounted for. Typically, MTT algorithms deal with missed detections and false alarms. Besides, MTT are designed to manage the so-called data association or measurement-origin uncertainty (MOU) problem, i.e., the fact that it is unknown which target (if any) generated a specific measurement. The MTT problem is further complicated by the presence of multiple sensors, as it is the case of multiple receivers of a MPRN. 

In the next section, we provide an overview of the implementation of the SPA-based MTT algorithm \cite{MeyBraWilHla:J17,MeyKroWilLauHlaBraWin:J18} relying on the use of particles, as well as how the MTT can benefit from the integration of the ARCE localization method. This integration, in fact, allows to perform a tailored sampling of the state space.

\section{MTT Algorithms for MPRNs}
\label{sec:MTT_algorithms}

\subsection{Objective and Challenges}
\label{subsec:objective-and-challenges}

MTT algorithms aim at sequentially estimating --- across multiple time
scans --- the \textit{states}, e.g., positions and velocities, of multiple targets by exploiting both the measurements generated by
multiple sensors and an a-priori knowledge on the target dynamics.
Let us denote with $\V{s}_{k,1},\ldots,\V{s}_{k,L}$ the unknown states of $L$ targets at time $k$, where $\V{s}_{k,\ell} \deq [ \V{x}_{k,\ell}^{\T} \,\, \V{v}_{k,\ell}^{\T} ]^{\T}$, and $\V{x}_{k,\ell}$ and $\V{v}_{k,\ell}$ are 3D position and 3D velocity, respectively, of the $\ell$-th target.\footnote{Higher order kinematics might be included in the target state, e.g., acceleration and jerk, depending on the modeling of the target dynamics.}
We consider an MPRN comprising a single transmitter and $S$ receivers with the receiver labeled $i = 1$ co-located with the transmitter as in Section~\ref{sec:Single_target_localization}. Unlike the previous section though, here we explicitly account for the presence of multiple measurements at each receiver. Specifically, receiver $i$ produces  $M^{(i)}_k \geqslant 0$
measurements at time $k$, due to both the presence of multiple targets and clutter,
and each target $\ell$ produces a measurement $\rho_{k,m}^{(i)}$ at receiver $i$ with probability $P_{\text{d}}^{(i)}$, and it is missed
with probability $1 - P_{\text{d}}^{(i)}$.
If the $m$-th measurement $\rho_{k,m}^{(i)}$
at receiver $i$ is generated by the $\ell$-th target then it is modeled (see eq. \eqref{eq:bistatic-range-noisy}) as
\begin{equation}\label{eq:bist_measurements_MTT}
\begin{split}
  \rho_{k,m}^{(i)}=  \| \V{x}_{k,\ell} - \V{t}_k \| + \| \V{x}_{k,\ell} - \V{r}_k^{(i)} \| + w^{(i)}_{k,m},
\end{split}
\end{equation}
where
$w_{k,m}^{(i)}$ are zero-mean
(usually Gaussian distributed) random variables independent across $k$, $i$, and $m$.
It is worth noting that the
measurement only depends on the target's position $\V{x}_{k,\ell}$ and not on its velocity $\V{v}_{k,\ell}$, which remain unobserved and can only be inferred if the target's dynamics is taken into account.

The presence of multiple targets and the availability of multiple measurements --- some of which might be clutter-generated (i.e., false alarms) --- is the cause of the MOU problem, i.e., the unknown association of measurements with targets, whose complexity scales exponentially with the number of targets, sensors, and measurements. Indeed, even considering a single sensor with no false alarms, and assuming that each target may generate \textit{at most} one measurement,\footnote{In general, when a target does not produce any measurement at a given receiver, it is considered a \textit{missed detected} at that receiver.} known as the point-target assumption \cite[Sec. 2.3]{BarWilTia:B11}, the number of possible associations between the $L$ targets and the $M_{k}^{(i)}$ measurements is $L! / (L - M_{k}^{(i)})!$.
As an example, consider a case with $L = 4$ targets and $M_{k}^{(i)} = 2$ measurements: the number of possible associations is $12$. Adding one more target and one more measurement, the number of associations becomes $60$. Clearly, the number of associations also increases if measurements may stem from false alarms~\cite{DezBar:TR21}.

Up to this point we have assumed that the number of targets, $L$, is time-invariant, either known or unknown.
For many tracking scenarios, however, this assumption does not hold.
Indeed, targets may enter the field-of-view of the sensors or, in other words, \textit{appear} in the tracking scenario;
because of this,
not-associated measurements are not necessarily false alarms, but they might be
determined by newly observed targets.
Likewise, targets may leave the coverage area, or \textit{disappear} from the tracking scenario, thus not generating any more measurements at the sensors.
In these cases, the number of targets $L_{k}$ needs to be modeled as time-variant and, if unknown, can be estimated alongside the target states.
Several approaches can be used to handle these appearance and disappearance phases, known as track formation or initialization, and track termination \cite[Sec. 3.3]{BarWilTia:B11}.

\subsection{State-of-the-art MTT Algorithms}
\label{subsec:state_MTT}
State-of-art MTT algorithms can be essentially divided into ``vector-type'' algorithms, which describe the target states and the measurements by random vectors, and ``set-type'' algorithms, which instead are based on random finite sets (RFSs)~\cite{Mah:B07}. RFSs are practical to model target appearance and disappearance in a Bayesian framework, and to handle complex and hybrid continuous/discrete distributions. 
The first category
includes the joint probabilistic data association (JPDA) \cite[Sec. 6.4]{BarWilTia:B11} and the joint integrated probabilistic data association (JIPDA)\cite{ChaMor:B11,MusEva:J04} filters, which address the MOU and the estimation problems assuming that each measurement is related to at most one target --- i.e., the point-target assumption --- and the posterior probability density function (pdf) of each target state is Gaussian.
Multiple hypothesis tracking (MHT) methods \cite{Rei:J79,ChoMorRei:J19}
use a deferred decision logic, that is,
decisions about target-measurement associations exploit
multiple measurements collected in more than one time
scan (i.e., within a reference time interval).
Therefore, a tree of potential track hypotheses for each candidate target is built, and only the branch representing the most likely target-measurement associations over the reference time interval is maintained and further propagated.

Popular examples of set-type algorithms are instead the probability hypothesis density (PHD) filter \cite{Mah:J03} and the cardinalized PHD (CPHD) filter \cite{Mah:J07,VoVoCan:J07}.
They both compute the posterior PHD of the multitarget state in a sequential fashion 
and the CPHD filter represents a generalization of the PHD filter which additionally propagates the cardinality distribution of the RFS.
The iterated-corrector (C)PHD 
filter \cite{NagCla:C11} and the partition-based (C)PHD 
filter \cite{NanBloCoaRab:J16} represent computationally
feasible multisensor extension of the (C)PHD methods \cite{BraMarMatWil:J13}.
Multi-Bernoulli (MB) filters approximate the posterior multitarget state RFS
by an MB RFS, or by a mixture of MB RFSs
\cite{Mah:B07}, where each MB RFS component 
corresponds to a global target-measurement association hypothesis.  
Labeled RFS-based multitarget tracking methods, such as the $\delta$-generalized labeled MB ($\delta$-GLMB) filter~\cite{VoVo:J13} and the labeled MB (LMB) filter~\cite{ReuVoVoDie:J14}, augment target states introducing distinct labels in order to maintain track continuity. 

Despite the wide menu of state-of-the-art MTT algorithms of either type, i.e., vector-type and set-type, their complexity usually does not scale well in large MTT scenarios.

\subsection{SPA-based Multisensor MTT Algorithm} \label{sec:multisensor_multitarget_tracking_SPA}
The issue of computational complexity and scalability of state-of-the-art MTT methods is well addressed by a recent and innovative particle-based Bayesian MTT approach, which
relies on the use of a factor graph and the SPA, i.e., the \emph{sum-product algorithm}~\cite{BraGagSolMenFerLepNicWilBraWin,MeyBraWilHla:J17,MeyKroWilLauHlaBraWin:J18,SolMeyBraHla:J19,GagSolMeyHlaBraFarWin:J20,GagBraSolMeyHlaMoe:J22}. The factor graph is used to represent the statistical dependencies among the random variables of the MTT model, while the Bayesian inference
is efficiently and reliably approximated by the SPA.
This technique is able to exploit conditional independence properties of random variables to achieve a drastic reduction of the computational complexity, handling efficiently both the data association and the fusion of measurements from multiple receivers --- even heterogeneous \cite{GagBraSolMeyHlaMoe:J22}. In this respect, the SPA enables an efficient calculation of association probabilities for \textit{soft}\footnote{A single-sensor MTT algorithm that uses a \textit{soft} data association technique does not select a specific measurement to update a target's state; it rather updates the target's state by averaging over all possible target-measurement combinations suitably weighted by their association probabilities.
Conversely, with a \textit{hard} data association technique a single-sensor MTT algorithm updates a target's state with a single measurement, selected as the
one maximizing the association probability \cite[Sec. 2.4]{BarWilTia:B11}.} target-measurement associations. For this reason the
SPA-based MTT method is particularly suitable for large-scale MPRNs tracking scenarios involving a large number of targets, receivers, and measurements, and enabling its use on resource-limited devices. 

To account for the estimation of both the number of targets and their states, the state of each target $\V{s}_{k,\ell}$ is augmented by a Bernoulli random variable $r_{k,\ell}$ that is equal to $1$ if the target is present, and $0$ otherwise; consequently,
$L_{k}$ represents the number of \textit{potential} or tentative targets.
The Bayesian inference about the
presence and the state of potential target $\ell$ at time $k$ is then based on the joint posterior pdf $f (\V{s}_{k,\ell}, r_{k,\ell} | \V{\rho}_{1:k})$, where $\V{\rho}_{1:k}$ is the vector comprising all the measurements from all the receivers since the initial time up to the current time $k$.
Specifically, the existence of potential target $\ell$ is confirmed if the marginal posterior probability mass function (pmf) $p (r_{k,\ell} = 1 | \V{\rho}_{1:k})$ is above a prefixed threshold\footnote{The \textit{estimated} number of targets is the cardinality of the set $\{\ell \, : \, p(r_{k,\ell} = 1 | \V{\rho}_{1:k}) > P_{\text{th}} \}$.} $P_{\text{th}}$~\cite[Ch. 2]{Poo:B94}, and an estimate of the potential target's state is obtained from the marginal posterior pdf $f (\V{s}_{k,\ell} | r_{k,\ell} = 1, \V{\rho}_{1:k})$ through,
for example, the minimum mean square error estimator (MMSE)~\cite[Ch. 4]{Poo:B94}. Note that these marginal posterior pdfs/pmfs can be obtained from the joint posterior pdf above by simple elementary operations, including marginalization.
The SPA-based MTT algorithm
computes an approximated version of the joint posterior pdf $f (\V{s}_{k,\ell}, r_{k,\ell} | \V{\rho}_{1:k})$ --- called \textit{belief} --- for all
the potential targets by employing an iterative version of the SPA on a suitably devised factor graph~\cite{MeyKroWilLauHlaBraWin:J18};
for future reference, we refer to the belief approximating the joint posterior pdf for potential target $\ell$ at time $k$ as $\tilde{f}(\V{s}_{k,\ell},r_{k,\ell})$.
The complexity of
the SPA-based MTT algorithm scales only quadratically in the number of potential targets, linearly in the number of transmitter-receiver pairs, and linearly in the number of measurements per receiver. Moreover, it  outperforms previously proposed methods in terms of accuracy ~\cite{MeyBraWilHla:J17,MeyKroWilLauHlaBraWin:J18,GagSolMeyHlaBraFarWin:J20}. 
Finally, since the SPA-based MTT method uses a particle-based implementation,
it is potentially suitable for arbitrary non-linear and non-Gaussian problems~
\cite{BraGagSolMenFerLepNicWilBraWin,MeyBraWilHla:J17,MeyKroWilLauHlaBraWin:J18,SolMeyBraHla:J19,GagSolMeyHlaBraFarWin:J20,GagBraSolMeyHlaMoe:J22}.

In the following section we illustrate how the ARCE localization method described in Section~\ref{subsec:const_sing_tar_localization} can be efficiently combined with the SPA-based MTT approach.

\subsection{Combination of the ARCE Localization with the SPA-based MTT Algorithm} \label{sec:ARCE_plus_SPA}

The SPA-based MTT algorithm has shown its advantages in terms of both accuracy and computational complexity compared to alternative approaches. Nonetheless, its computational burden can still rapidly grow in a 3D scenario as the one considered here, because of the high number of particles required to effectively sample the 6D potential target state space.
The use of a limited number of particles is thus desired, which, however, 
can lead to particle degeneracy and impoverishment,\footnote{\textit{Degeneracy} occurs when, over time, most of the weight of the entire set of particles is concentrated on few particles, whereas the remaining particles have a negligible weight. This effect is generally addressed through resampling, which however might cause particle \textit{impoverishment}, that is, a reduction of
particle diversity \cite{LiBolDju:J15}.} and more generally to an inaccurate representation of the pdfs/beliefs.
This is particularly relevant in our range-only sensing context, when initializing the state of a newly observed target from a bistatic measurement at the single receiver node. Indeed, the lack of any angle information requires the prior pdf of the potential target state, in particular the component related to the 3D position, to cover a large volume, potentially the entire
focaloid induced by the bistatic range measurement as well as transmitter and receiver positions;\footnote{A \textit{focaloid} is a shell bounded by two confocal ellipsoids; it reduces to a \textit{spherical shell} when transmitter and receiver are colocated.} clearly, the resulting prior pdf cannot be reliable represented by a small number of particles. Needless to say, both an inaccurate prior distribution choice and its rough representation can propagate via the target dynamics over time, degrading the overall tracking performance.

Inspired by the rationale behind the ARCE localization, here the aim is to
capitalize on some prior angular information, related to the potential target
position, in order to mitigate
the negative effects caused by the use of a limited number of particles, and thus improve
the
tracking performance.
The angular side information can be acquired either through physical considerations,
e.g., the antenna beamwidth of the transmitter as in the plain ARCE strategy, or leveraging the knowledge about the predicted
distribution of the potential target state.
We propose an efficient methodology to embed the ARCE location estimate 
within the SPA-based MTT algorithm and thus enable a smarter sampling of the potential target state space;
this process is driven by the ARCE location estimate, which is
essentially memoryless, i.e., unaffected by the past, and mainly depends on the measurements available at current time.

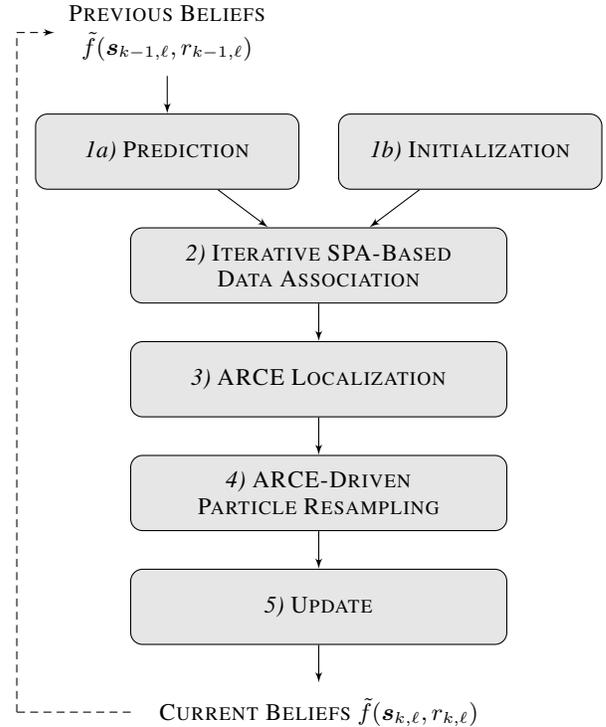
\begin{figure}[!t]
    \centering%
    \begin{tikzpicture}[auto,>=latex']
    
        \node [int] (b) {\textsc{\textit{2)} Iterative SPA-Based} \\ \textsc{Data Association}};
        \node [int] (c) [below=.5cm of b] {\textsc{\textit{3)} ARCE Localization}};
        \node [int] (d) [below=.5cm of c] {\textsc{\textit{4)} ARCE-Driven} \\ \textsc{Particle Resampling}};
        \node [int] (e) [below=.5cm of d] {\textsc{\textit{5)} Update}};
        
		\node [int2] (a1) [above left=0.5cm and -2.25cm of b] {\textsc{\textit{1a)} Prediction}};
		\node [int2] (a2) [above right=0.5cm and -2.25cm of b] {\textsc{\textit{1b)} Initialization}};
		\node [inner sep=2mm, font=\small, align = center] (bea) [above = 0.5cm of a1] {\textsc{Previous Beliefs} \\[1mm] $\tilde{f}(\V{s}_{k-1,\ell},r_{k-1,\ell})$};
		
        \node [inner sep=2mm, font=\small, align = center] (f) [below=0.5cm of e] {\textsc{Current Beliefs $\tilde{f}(\V{s}_{k,\ell},r_{k,\ell})$}};
        
        \path[->] (bea) edge node {} (a1);
		\path[->] (a1) edge node {} (b);
		\path[->] (a2) edge node {} (b);
		\path[->] (b) edge node {} (c);
		\path[->] (c) edge node {} (d);
		\path[->] (d) edge node {} (e);
		\path[->] (e) edge node {} (f);
		
	    \coordinate (l1) at ($(f)+(-2.5cm,0)$);
	    \coordinate (l3) at ($(a1)+(-2cm,0)$);
	    \coordinate (l5) at ($(bea)+(-1.5cm,0)$);
	    \coordinate (l2) at (l3|-l1);
	    \coordinate (l4) at (l3|-l5);
	    
		\draw[densely dashed,-] (l1) to (l2);
		\draw[densely dashed,-] (l2) to (l3);
		\draw[densely dashed,-] (l3) to (l4);
		\draw[densely dashed,->] (l4) to (l5);
		
    \end{tikzpicture}
    \caption{Block diagram
        reporting the steps of the proposed ARCE-enhanced SPA-based MTT algorithm performed at time $k$.
        }
    \label{fig:proposed-method-diagram}
\end{figure}%

Figure~\ref{fig:proposed-method-diagram} shows the steps of the proposed ARCE-enhanced SPA-based MTT algorithm, already briefly discussed in Fig.~\ref{fig:notional-sketch}.
The beliefs computed at the previous time
scan $k-1$, representative of the
potential targets observed so far, are predicted to current time $k$ by means of a kinematic model.
Meanwhile, new
potential target states are initialized so that newly observed targets, i.e.,
newly-appearing targets, are promptly tracked. 
Ideally, this initialization should involve the measurements collected by all the receivers at current time $k$, procedure that
demands a high computational cost. 
As an example, let us consider $S = 2$ receivers each with a single measurement, that is, $M_{k}^{(i)} = 1$ for $i = 1,2$.
Both measurements can be false alarms or be generated by the same newly observed target; or each measurement can be generated by different newly observed targets; or the measurement from the first receiver can be generated by a newly observed target while the other be a false alarm, and vice versa.
As seen, even in this simple case with only two measurements from two receivers, the initialization step should
account for five different scenarios.
Therefore, in order to limit the complexity, only measurements from one of the receivers are considered for the initialization step; specifically, the
$M_{k}^{(1)}$ measurements collected by the monostatic active radar (i.e., the receiver labeled $i = 1$ co-located with the transmitter),
since this sensor is deemed more reliable, in terms of detectability, compared to the other passive receivers.
Then, the iterative SPA-based data association procedure
computes the \textit{soft} association probabilities for each potential target-measurement combination.
These association probabilities are used as they are in the update step, according to the common SPA-based MTT framework \cite{MeyKroWilLauHlaBraWin:J18}, and are transformed into \textit{hard} potential target-measurement associations in order to cluster the measurements and accomplish single-snapshot ARCE localization based
on each group.
Additionally, ARCE localization algorithm exploits some prior angular information to compute a potential target's position estimate.
Two approaches are herein pursued, based on how this information is acquired.
The \textit{non-adaptive} (NAD) approach uses the physical beamwidth and looking direction of the active radar antenna to establish the angular constraints; hence, these constraints are time-invariant and equal for all
the potential targets.
The \textit{adaptive} (AD) countepart exploits an appropriate \textit{virtual beam} to define bespoke angular constraints in the ARCE process.
This
virtual beam is
unique for each
potential target, and is given by the intersection of the active antenna beam and a tailored beam: the latter
points towards the potential target's predicted position, and its beamwidth is proportional to the uncertainty of such predicted position in the angular domain.

The last two steps are key ingredients of the proposed method. They refer to the ARCE-driven particle resampling, used to obtain a smarter sampling of the potential target state space based on the ARCE localization estimates, and the update step used to eventually obtain the beliefs at current time.
Hereafter, a detailed description of each step performed at time $k$ is provided.

\subsubsection{Prediction and Initialization}
\label{sec:pred_step}
The input to the prediction step is the set of ${L}_{k-1}$ previous beliefs
$\tilde{f}(\V{s}_{k-1,\ell},r_{k-1,\ell})$, $\ell \in \{ 1,\ldots,L_{k-1} \}$, representing the joint posterior pdfs $f(\V{s}_{k-1,\ell}, \linebreak r_{k-1,\ell} | \V{\rho}_{1:k-1})$ computed at time $k - 1$.
Following the derivation in \cite[Sec. VI]{MeyBraWilHla:J17}, the previous belief of potential target $\ell$ for $r_{k-1,\ell} = 1$, i.e., $\tilde{f}(\V{s}_{k-1,\ell},r_{k-1,\ell} = 1)$, is represented by a set of $N_{\text{p}}$ weighted particles\footnote{The notation $i|j$ as subscript indicates a random variable/vector evaluated at time $i$ given the measurements from the initial time up to time $j$.} $\{ \V{s}_{k-1|k-1,\ell}^{(p)}, \omega_{k-1|k-1,\ell}^{(p)} \}_{p=1}^{N_{\text{p}}}$, whose weights, contrary to conventional particle filtering \cite{AruMasGorCla:J02}, do not sum to one. Indeed, it is straightforward to verify that $p_{k-1|k-1,\ell}^{\text{e}} \deq \sum_{p = 1}^{N_{\text{p}}} \omega_{k-1|k-1,\ell}^{(p)}$ is approximately equal to $p(r_{k-1,\ell} = 1 | \V{\rho}_{1:k-1})$, i.e., the posterior probability of existence of potential target $\ell$ at time $k - 1$.
During
the prediction step,
this set of weighted particles
is converted into a new set of weighted particles $\{ \V{s}_{k|k-1,\ell}^{(p)}, \omega_{k|k-1,\ell}^{(p)} \}_{p=1}^{N_{\text{p}}}$, 
that approximates the joint predicted pdf $f(\V{s}_{k,\ell}, r_{k,\ell} | \V{\rho}_{1:k-1})$. Specifically,  $\omega_{k|k-1,\ell}^{(p)}=p^{\text{s}}\omega_{k-1|k-1,\ell}^{(p)}$, where $p^{\text{s}}$ is the target
survival probability.
The particle evolution is achieved by utilizing an appropriate 
kinematic model, described by the transition pdf $f(\V{s}_{k,\ell} | \V{s}_{k-1,\ell})$.

Meanwhile, as mentioned above,
new potential target states are initialized to account for newly observed targets.
First, let us recall that, in order to limit the computational cost, only the $M_{k}^{(1)}$ measurements produced by the monostatic active radar are used in the initialization step;
therefore, in order to account for newly observed targets, a set of weighted particles $\{ \V{s}_{k|k-1,m'}^{(p)}, \omega_{k|k-1,m'}^{(p)} \}_{p=1}^{N_{\text{p}}}$, with $m' = L_{k - 1} + m$, is added\footnote{The subscript $k | k-1$ is kept for consistency with the notation used for the predicted sets of particles. Clearly, new potential targets are independent of the previous time scan $k-1$.} for each measurement $m \in \{ 1,\ldots,M_{k}^{(1)} \}$. 
The 3D position component of each particle, i.e., $\V{x}_{k|k-1,m'}^{(p)}$, is drawn from a distribution --- usually Gaussian, according to the measurement model in eq. \eqref{eq:bist_measurements_MTT} --- with mean the range measurement $\rho_{k,m}^{(1)}$ converted into Cartesian coordinates assuming an angle uniformly distributed within the transmitter's antenna beam, and standard deviation in accordance to the noise $w_{k,m}^{(1)}$ in eq. \eqref{eq:bist_measurements_MTT};
the 3D velocity component, i.e., $\V{v}_{k|k-1,m'}^{(p)}$, is drawn from a
Gaussian distribution independent of $k$, $m'$, and $p$, with mean zero and scalar
covariance matrix whose non-zero element is related to the target's maximum speed, according to the one-point initialization provided in \cite[Sec. 3.2.2]{BarWilTia:B11}. Finally, homogeneous particle weights are set, i.e., $\omega_{k | k-1,m'}^{(p)}=\frac{p^{\text{b}}}{N_\text{p}}$, with $p^{\text{b}} \ll 1$ the assumed birth probability.
It is worth noting that by using this mechanism, the number of potential targets --- i.e., of particle sets ---  grows indefinitely over time; indeed, following the initialization, the number of potential targets at time $k$ becomes $L_{k} \deq L_{k - 1} + M_{k}^{(1)}$.
Therefore, in order to keep a tractable computational complexity, a pruning step is performed at each time
scan $k$, before prediction and initialization, in order to remove all potential targets whose probability of existence is below a prefixed threshold \cite{BraGagSolMenFerLepNicWilBraWin,MeyKroWilLauHlaBraWin:J18}.

\subsubsection{Iterative SPA-Based Data Association}
\label{sec:data_association_step}

The $L_k$ sets of weighted particles obtained at the previous step,
and all
measurements collected by all receivers at time $k$ are used to compute the soft association probabilities for each potential target-measurement combination
according to the SPA-based data association
algorithm as described in~\cite{MeyKroWilLauHlaBraWin:J18,GagSolMeyHlaBraFarWin:J20}.
These soft association probabilities are used
as they are in the update step, whereas they are transformed into hard potential target-measurement associations so as to cluster the measurements into groups to be used in the next ARCE localization step.
The hard potential target-measurement associations are obtained by applying a maximum-a-posteriori criterion to the approximated measurement-oriented data association pmfs,\footnote{The measurement-oriented data association pmfs encode the probabilities of each measurement being either generated  by a potential target $\ell$ or being a false alarm.} computed as described in \cite[Sec. VI-B]{MeyKroWilLauHlaBraWin:J18}. 

\subsubsection{ARCE Localization}
\label{sec:ARCE_loc_step}
During this step an estimate of each potential target position, denoted by $\V{x}_{k,\ell}^{\text{ARCE}}$, is obtained using the ARCE localization algorithm described in Section~\ref{subsec:const_sing_tar_localization}.
As clearly shown also in Fig.~\ref{fig:block-diagram}, in order to compute $\V{x}_{k,\ell}^{\text{ARCE}}$ the ARCE localization process requires as input a set of bistatic-range
measurements associated to potential target $\ell$, as well as specific angular constraints.
The set of
measurements is obtained through
the hard potential target-measurement
associations
computed at the previous step.
The angular constraints are selected according to two different approaches. 
When using the NAD --- non-adaptive --- approach, the angular constraints just reflect the physical beam of the transmitter antenna; therefore, they are the same for all the potential targets.
The AD --- adaptive --- approach, instead, defines different angular constraints for each potential target $\ell$ according to a bespoke virtual beam, obtained as the intersection of the physical beam of the transmitter antenna and a tailored beam.
This tailored beam is steered towards the predicted potential target position obtained as the weighted sum (according to $\omega_{k | k-1, \ell}^{(p)}$) of the particles $\V{x}_{k|k-1,\ell}^{(p)}$.To compute the width of the
tailored beam in azimuth (XY-plane) and elevation (XZ-plane), instead,
first the particles $\V{x}_{k|k-1,\ell}^{(p)}$ are converted from Cartesian coordinates to spherical coordinates; then, the standard deviations of azimuth and elevation, denoted by $\sigma^{\,\text{az}}_{k,\ell}$ and $\sigma^{\,\text{el}}_{k,\ell}$, respectively, are computed to measure the potential target predicted uncertainty along
the principal planes.
Finally, the widths in azimuth and elevation are set to, respectively, $d^{\,\text{az}}_{k,\ell} =  2 \tilde{C} \sigma^{\,\text{az}}_{k,\ell}$ and $d^{\,\text{el}}_{k,\ell} = 2 \tilde{C} \sigma^{\,\text{el}}_{k,\ell}$, where $\tilde{C}$ is a scaling factor used to widen ($\tilde{C} > 1$) or narrow ($\tilde{C} < 1$) the tailored beam.

\subsubsection{ARCE-Driven Particle Resampling}
\label{sec:resampling_step}
Objective of this step is to provide a more accurate/reliable sampling of the potential target state space, or, equivalently, a more accurate representation of the potential target belief, exploiting the position estimate $\V{x}_{k,\ell}^{\text{ARCE}}$ provided by the ARCE.
The idea comes from the consideration that the number of particles ---
limited to keep a tractable computational complexity --- might not be enough to well describe the potential target belief.
In addition, this coarse representation can  propagate over time, eventually leading to the particle impoverishment and a performance degradation. Hence,
to prevent
such impairments, the intuition is to replace the less-significant
particles representing the predicted potential target position, i.e., $\V{x}_{k|k-1,\ell}^{(p)}$, with new particles drawn from a suitable distribution centered in the ARCE
localization estimate. The aforementioned distribution, referred to as ARCE-based distribution, is a Gaussian with mean $\V{x}_{k,\ell}^{\text{ARCE}}$ and prefixed standard deviation $\sigma^{\text{ARCE}}$ used to model the uncertainty of the estimated ARCE location. Below a detailed description of the substitution procedure is provided.

For each potential target $\ell$,
let us assume
without loss of generality that the weights $\omega_{k | k-1,\ell}^{(p)}$ are ordered from the smallest to the largest, i.e., $\omega_{k | k-1,\ell}^{(p)} \leqslant \omega_{k | k-1,\ell}^{(q)}$ for $p < q$.
Then, let us denote with $\Set{P} \deq \{ 1, \ldots, N_{\text{g}} \}$ the set of indices representing the fraction $1 - \alpha_{\text{r}}$, $\alpha_{\text{r}} \in (0,1)$, of less significant particles that will be replaced; $N_{\text{g}}$ is therefore the largest
value in $\{ 1, \ldots, N_{\text{p}} \}$ such that the following condition holds true:\footnote{Note that if the weights are uniform --- as for the new potential targets --- this procedure is equivalent to a random selection of the particles to replace. Specifically, each particle $p$ is replaced with probability $1 - \alpha_{\text{r}}$ and maintained with probability $\alpha_{\text{r}}$.}
\begin{align}
	\dfrac{\sum_{q = 1}^{N_{\text{g}}} \omega_{k | k-1,\ell}^{(q)}}{\sum_{p = 1}^{N_{\text{p}}} \omega_{k | k-1,\ell}^{(p)}} \leqslant (1 - \alpha_{\text{r}})
\end{align}
The ARCE-driven set of weighted particles, denoted by $\{ \overline{\V{s}}_{k|k-1,\ell}^{(p)}, \overline{\omega}_{k|k-1,\ell}^{(p)} \}_{p = 1}^{N_{\text{p}}}$, is 
built as follows. The particle $\overline{\V{s}}_{k|k-1,\ell}^{(p)}$ is
\begin{align}
	\overline{\V{s}}_{k|k-1,\ell}^{(p)} =
	\begin{dcases}
		\Big[ \check{\V{x}}_{k|k-1,\ell}^{(p)\T} \,\, \V{v}_{k|k-1,\ell}^{(p)\T} \Big]^{\T} \, ,	&	p \in \Set{P} \, , \\[1mm]
		\V{s}_{k|k-1,\ell}^{(p)}\, ,	&	p \notin \Set{P} \, ,
	\end{dcases}
\end{align}
where $\check{\V{x}}_{k|k-1,\ell}^{(p)}$ is drawn from the ARCE-based distribution; note that only the 3D position component of the particle is replaced if $p \in \Set{P}$, whereas the 3D velocity component $\V{v}_{k|k-1,\ell}^{(p)}$ is kept since the ARCE localization algorithm does not provide any velocity information.
The weight $\overline{\omega}_{k|k-1,\ell}^{(p)}$ is
\begin{align}
	\overline{\omega}_{k|k-1,\ell}^{(p)} = \Bigg( \sum_{d = 1}^{N_{\text{p}}}\omega_{k|k-1,\ell}^{(d)} \Bigg) \times
	\begin{dcases}
		\dfrac{1 - \alpha_{\text{r}}}{N_{\text{g}}} \, ,	&	p \in \Set{P}\, , \\[1mm]
		\dfrac{\alpha_{\text{r}} \, \omega_{k|k-1,\ell}^{(p)}}{\sum_{q \notin \Set{P}} \omega_{k|k-1,\ell}^{(q)}} \, ,	&	p \notin \Set{P} \, .
	\end{dcases}
\end{align}
We note that one could also consider to calculate the weight  $\overline{\omega}_{k|k-1,\ell}^{(p)}$ according to the standard  sequential importance sampling (SIS) (cf. 
Algorithm 2 in \cite{CappeGM07_95}). However, this approach is numerically unstable, since the weight $\overline{\omega}_{k|k-1,\ell}^{(p)}$  of the particle $\overline{\V{s}}_{k|k-1,\ell}^{(p)}$ drawn from the ARCE-based distribution, for $p\in\Set{P}$, can result zero if the particle $\overline{\V{s}}_{k|k-1,\ell}^{(p)}$ is not compatible with the transition pdf $f(\V{s}_{k,\ell} | \V{s}_{k-1,\ell})$ that models the target dynamic. Then, the devised method 
addresses this numerical instability by ensuring that \textit{i)} the weights of the substituted particles are uniform and retain a fraction $1 - \alpha_{\text{r}}$ of the total weight of the set; \textit{ii)} the weights of the remaining particles are unchanged except that for a normalization factor that lets them retain a fraction $\alpha_{\text{r}}$ of the total weight of the set; and \textit{iii)} the sum of all the weights is unchanged, that is, $\sum_{p = 1}^{N_{\text{p}}} \omega_{k | k-1,\ell}^{(p)} = \sum_{q = 1}^{N_{\text{p}}} \overline{\omega}_{k | k-1,\ell}^{(q)}$.

\begin{figure*}[!t]

	\centering
	\subfloat[\textit{Non-adaptive} (NAD) Approach - XY-Plane]{
		\includegraphics{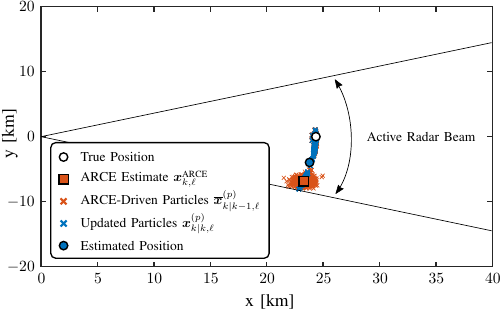}
		\label{fig:nad-xy} 
	} \hspace{2mm}	
	\centering
	\subfloat[\textit{Non-adaptive} (NAD) Approach - XZ-Plane]{
		\includegraphics{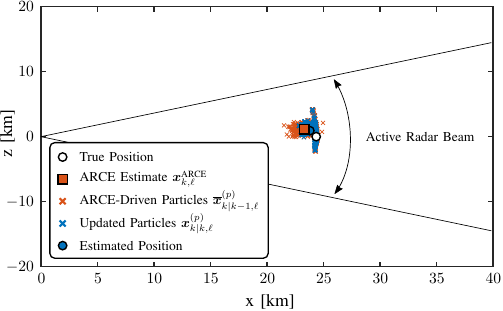}
		\label{fig:nad-xz} 
	}	

	\centering
	\subfloat[\textit{Adaptive} (AD) Approach - XY-Plane]{
		\includegraphics{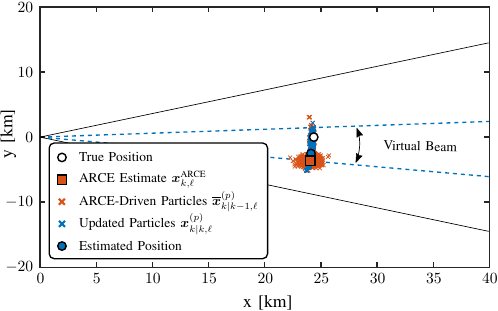}
		\label{fig:ad-xy} 
	} \hspace{2mm}	
	\centering
	\subfloat[\textit{Adaptive} (AD) Approach - XZ-Plane]{
		\includegraphics{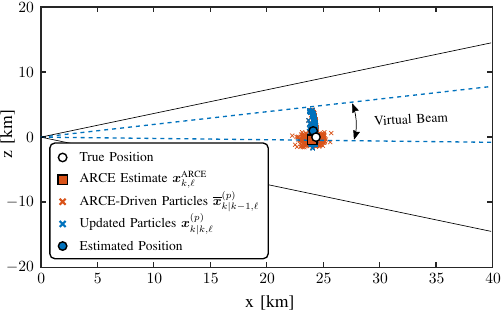}
		\label{fig:ad-xz} 
	}	
      
		\caption{Illustrations of the proposed ARCE-enhanced SPA-based MTT algorithm --- both the NAD (top plots) and the AD (bottom graphs) approaches --- in a 3D single target
				scenario, assuming the active radar located at $[0 \,\, 0 \,\, 0]^{\T}$
				whose antenna is pointing towards the X-axis;
				the plots show the XY-planes (left panels) and the XZ-planes (right panels).
				The NAD approach considers the active radar beamwidth to establish the angular constraints used
				for the computation of the ARCE
				estimate;
				the
				AD approach, instead, utilizes 
				virtual beams as described in Section~\ref{sec:ARCE_loc_step}.
						}
	\vspace{-2mm}
	\label{fig:nad-ad}
\end{figure*}

We note that at the end of this step the
a-priori knowledge of the monostatic active radar
beam can be also used
to
penalize the
particles lying outside the constrained region
by applying an acceptance/rejection process as described in~\cite{ShaoHL:J10}.

\subsubsection{Update}
\label{sec:meas_update}

According to the implementation of the SPA-based MTT algorithm
in \cite[Sec. VI]{MeyBraWilHla:J17}, the ARCE-driven sets of weighted particles $\{ \overline{\V{s}}_{k|k-1,\ell}^{(p)}, \overline{\omega}_{k|k-1,\ell}^{(p)} \}_{p = 1}^{N_{\text{p}}}$, $\ell \in \{ 1, \ldots, L_{k} \}$,
are updated using
the measurements collected by all the receivers at current time $k$ and the soft association probabilities computed
at the iterative SPA-based data association step.
The updated sets of weighted particles, denoted $\{ \V{s}_{k|k,\ell}^{(p)}, \omega_{k|k,\ell}^{(p)} \}_{p = 1}^{N_{\text{p}}}$,
represent the beliefs of the potential targets at current time, i.e., $\tilde{f}(\V{s}_{k,\ell},r_{k,\ell} = 1)$, which in turn approximate the joint posterior pdfs $f(\V{s}_{k,\ell},r_{k,\ell} = 1 | \V{\rho}_{1:k})$.
We recall from Section~\ref{sec:multisensor_multitarget_tracking_SPA} that potential target $\ell$ is confirmed if the marginal posterior pmf $p(r_{k,\ell} = 1 | \V{\rho}_{1:k}) \approx p_{k|k,\ell}^{\text{e}}$ is above the threshold $P_{\text{th}}$, and that an estimate of its state is obtained from the marginal posterior pdf $f(\V{s}_{k,\ell} | r_{k,\ell} = 1, \V{\rho}_{1:k}) \approx \tilde{f}(\V{s}_{k,\ell},r_{k,\ell} = 1) / p_{k|k,\ell}^{\text{e}}$.

Following the update, a resampling of the particles may be required
to mitigate degeneracy \cite{LiBolDju:J15}; this process results in the weights $\omega_{k|k,\ell}^{(p)}$ to be all equal and whose sum is the posterior probability of existence $p_{k|k,\ell}^{\text{e}}$.

Figure~\ref{fig:nad-ad} provides illustrations of the proposed ARCE-enhanced SPA-based MTT algorithm in a 3D single target scenario at a generic time
scan $k$, assuming that the active radar is located at $\V{t}_{k} = [0 \,\, 0 \,\, 0]^{\T}$
with its antenna pointing towards the X-axis; the left-hand side plots (panels \subref{fig:nad-xy} and \subref{fig:ad-xy}) and the right-hand side plots (panels \subref{fig:nad-xz} and \subref{fig:ad-xz}) show the projections of all the 3D points onto, respectively, the XY- and the XZ-planes.
The top plots refer to the NAD case, that is, when the angular constraints used within the ARCE localization algorithm (cf. Sec.~\ref{sec:ARCE_loc_step}) coincide with the physical beam of the transmitter's antenna.
All figures show the true position of the target (white circle), the ARCE estimate (red square), the set of ARCE-driven particles computed as described in Section~\ref{sec:resampling_step} (red crosses), the set of updated particles obtained as illustrated in Section~\ref{sec:meas_update} (blue crosses), and the final estimated position of the target (blue circle).
As expected, the ARCE estimate is within the active radar beam and, especially looking at the XZ-plane in Fig.~\ref{fig:nad-xz}, close to the true target position, allowing a better sampling of the state space in this relevant region.
The bottom figures show the same example when the AD approach (exploiting a virtual beam) is adopted.
The ARCE estimate is now restricted to the virtual beam, designed as described above; this avoids to spread the new $N_{\text{g}}$ particles, generated during the
ARCE-driven particle resampling
step in a region where it is less likely to observe the target, as happens, for example, with the NAD approach in Fig.~\ref{fig:nad-xy}.

\begin{figure*}[!t]
	\centering
	\subfloat[Target 1 - XY-Plane]{
		\includegraphics{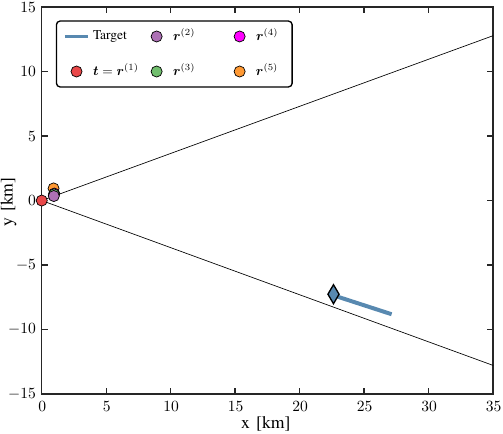}
		\label{fig:simulated_trajectories-border-xy} 
	} \hspace{2mm}	
	\centering
	\subfloat[Target 1 - XZ-Plane]{
		\includegraphics{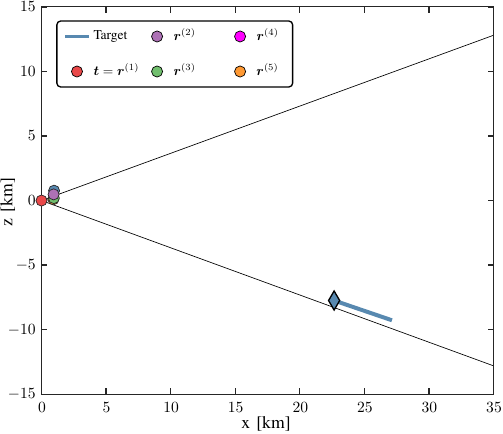}
		\label{fig:simulated_trajectories-border-xz} 
	}	

	\centering
	\subfloat[Target 2 - XY-Plane]{
		\includegraphics{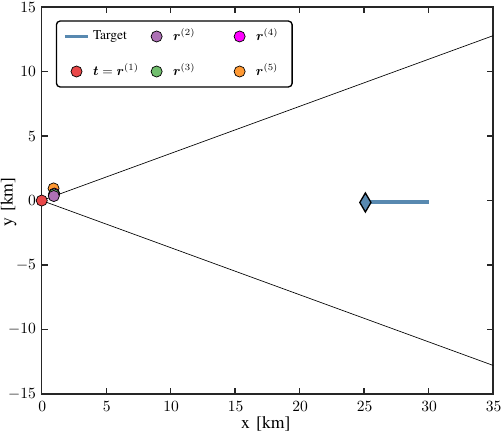}
		\label{fig:simulated_trajectories-centre-xy} 
	} \hspace{2mm}	
	\centering
	\subfloat[Target 2 - XZ-Plane]{
		\includegraphics{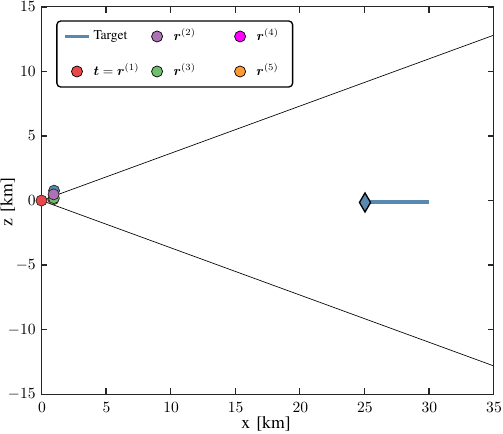}
		\label{fig:simulated_trajectories-centre-xz} 
	}
	\caption{Illustrations of the 3D simulated
		scenarios, with the active radar located at $\V{t}=\V{r}^{(1)}= [0 \,\, 0 \,\, 0]^{\T}$
		whose antenna is pointing towards the X-axis;
		the plots show the simulated trajectories of both target 1 (top panels) and target 2 (bottom panels) projected on the XY-plane (left panels) and the XZ-plane (right panels).
				}
			\label{fig:simulated_trajectories}
\end{figure*}

\begin{figure}[!t]
	\centering
	\subfloat[$\overline{\text{SNR}} = -20$ dB.]{
		\includegraphics{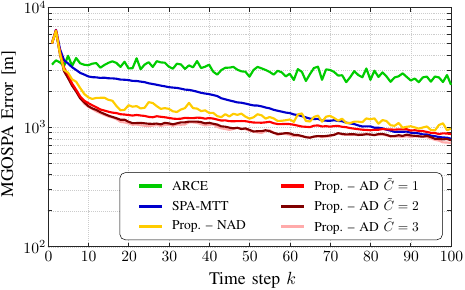}
		\label{fig:comp-ideal-scenario-border-pos_-20db}
	}
	
	\centering
	\subfloat[$\overline{\text{SNR}} = -10$ dB.]{
		\includegraphics{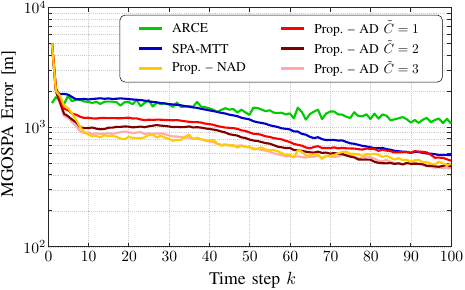}
		\label{fig:comp-ideal-scenario-border-pos_-10db}
	}
	
	\centering
	\subfloat[$\overline{\text{SNR}} = 0$ dB.]{
		\includegraphics{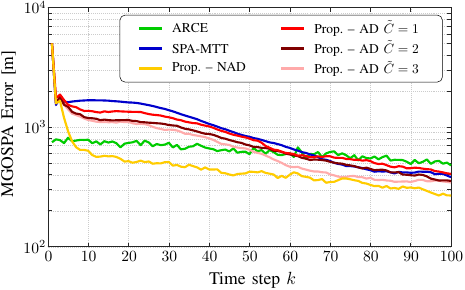}
		\label{fig:comp-ideal-scenario-border-pos_0db}
	}
	
	\caption{Comparison between the  ARCE localization algorithm (`ARCE'), the baseline SPA-based MTT algorithm (`SPA-MTT'), and the proposed algorithm (`Prop.') --- both NAD and AD --- in an ideal scenario ($P_{\text{d}} = 1$ and no false alarms) with target 1 moving close to the antenna's beam edge, in terms of MGOSPA error and on varying $\overline{\text{SNR}}$.}
	\label{fig:comp-ideal-scenario-border-pos}
\end{figure}

\begin{figure}[!t]
	\centering
	\subfloat[$\overline{\text{SNR}} = -20$ dB.]{
		\includegraphics{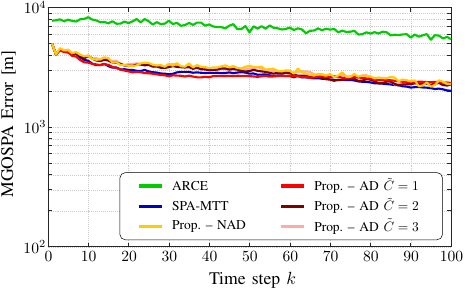}
		\label{fig:comp-ideal-scenario-central-pos_-20db}
	}
	
	\centering
	\subfloat[$\overline{\text{SNR}} = -10$ dB.]{
		\includegraphics{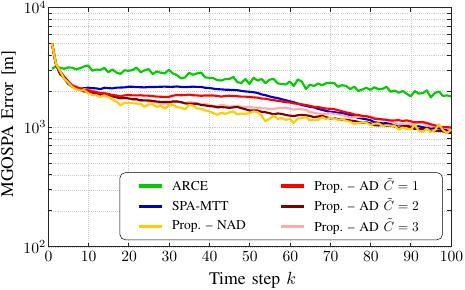}
		\label{fig:comp-ideal-scenario-central-pos_-10db}
	}
	
	\centering
	\subfloat[$\overline{\text{SNR}} = 0$ dB.]{
		\includegraphics{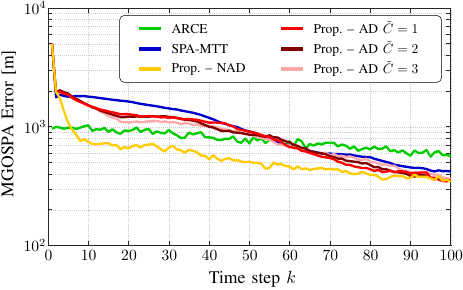}
		\label{fig:comp-ideal-scenario-central-pos_0db}
	}
	
	\caption{Comparison between the  ARCE localization algorithm (`ARCE'), the baseline SPA-based MTT algorithm (`SPA-MTT'), and the proposed algorithm (`Prop.') --- both NAD and AD --- in an ideal scenario ($P_{\text{d}} = 1$ and no false alarms) with target 2 moving in the middle of the antenna's beam, in terms of MGOSPA error and on varying $\overline{\text{SNR}}$.}
	\label{fig:comp-ideal-scenario-central-pos}
\end{figure}

\begin{figure}[!t]
		\centering
	\subfloat[Target 1 moving close to the antenna's beam edge.]{
		\includegraphics{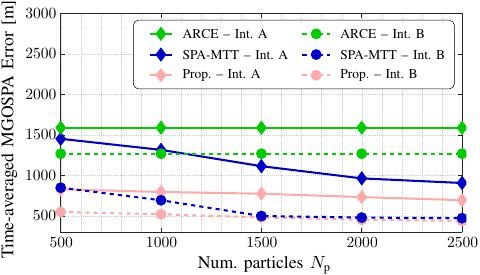}
		\label{fig:perf_init_final_tg1}
	}
	
		\centering
	\subfloat[Target 2 moving in the middle of the antenna's beam.]{
		\includegraphics{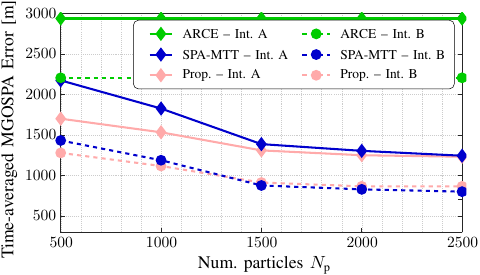}
		\label{fig:perf_init_final_tg2}
	}

	\caption{Comparison between the  ARCE localization algorithm (`ARCE'), the baseline SPA-based MTT algorithm (`SPA-MTT'), and the proposed algorithm (`Prop.') --- AD with $\tilde{C}=3$ --- in an ideal scenario ($P_{\text{d}} = 1$ and no false alarms), for
		$\overline{\text{SNR}}= -10$ dB, in terms of MGOSPA averaged over time \textit{interval A} (from scan 10 to scan 40, continuous lines) and time \textit{interval B} (from scan 41 to scan 100, dashed lines) and
		on varying the number of particles $N_\text{p}$.}
	\label{fig:perf_init_final}
\end{figure}

\section{Simulated experiments}
\label{sec:simulated_experiments}
In this section, the performance of the proposed algorithm, described in Section~\ref{sec:ARCE_plus_SPA}, is assessed via simulated experiments also in comparison with the baseline SPA-based MTT algorithm performing only target tracking~\cite{MeyBraWilHla:J17,MeyKroWilLauHlaBraWin:J18}.
 
\subsection{Simulation Setup}
\label{sec:sim_setup}

We simulate a 3D scenario with a stationary transmitter located at the origin of the reference system (i.e., $\V{t}_k = \V{t} = [0 \linebreak 0 \,\, 0]^\T$ km) and $S=5$ stationary receivers located, respectively, at $\V{r}^{(1)} = [0 \,\, 0 \,\, 0]^\T$ km, $\V{r}^{(2)} = [0.916 \,\, 0.941 \,\, 0.95]^\T$ km, $\V{r}^{(3)} = [0.973 0.541 \,\, 0.764]^\T$ km, $\V{r}^{(4)} = [0.955 \,\, 0.483 \,\, 0.191]^\T$ km, and $\V{r}^{(5)} = [0.936 \,\, 0.350 \,\, 0.477]^\T$ km. Note that the transmitter and receiver 1 are co-located (monostatic active radar), i.e., $ \V{t} = \V{r}^{(1)}$, and that the antenna
is steered towards the X-axis with half-beam width equal to 20 degrees.
We simulate two targets moving radially towards the active radar. In particular, target $1$ is moving close to the antenna's beam edge at an angle of -18 degrees in both azimuth and elevation, while target $2$ is moving in the middle of the antenna's beam; their initial range is either 30 or 35 km depending on the considered scenario as specified later. The trajectories of both target 1 and target 2 projected on the XY- and XZ- planes are shown in Fig.~\ref{fig:simulated_trajectories}.
Both targets are simulated for 100 time
scans with a scan time of 10 s, and their speed
is set to 5 m/s. The monostatic
and
bistatic measurements generated by the targets
are simulated according to eq.~\eqref{eq:bist_measurements_MTT} with $w_{k,m}^{(i)}\triangleq w^{(i)}$ being
distributed as a Gaussian random variable with mean 0 and standard deviation $\sigma_i = (B\sqrt{2SNR_i})^{-1}$, for $i = 1, \ldots, 5$. Here, $B = 20$ MHz represents the frequency bandwidth of the probing waveform,
and $SNR_i$ denotes the signal to noise ratio (SNR)
of the $i$-th transmitter/receiver pair that is function of the target's position \cite[eq. (27)]{AubBraDemMar:J22}, along with other parameters involved in the radar equation.

We compare the performance of the baseline SPA-based MTT algorithm performing only target tracking~\cite{MeyBraWilHla:J17,MeyKroWilLauHlaBraWin:J18} with the proposed method described in Section~\ref{sec:ARCE_plus_SPA}. In particular, for the proposed method we consider the NAD version
and the AD version for three values of $\tilde{C} = 1,2,3$ (cf. Sec.~\ref{sec:ARCE_loc_step}).
The performance of the different methods is measured according the generalized optimal sub-pattern assignment (GOSPA) metric \cite{RahGarSven:C17} that accounts for localization errors for correctly confirmed targets, as well as errors for missed targets and false targets; all the results are averaged over 200 Monte Carlo runs.
We simulate an ideal scenario without missed detections and false alarms (cf. Sec.~\ref{sec:sim_results_ideal}) as well as a more challenging scenario, where the detection probabilities of the receivers are lower than one, i.e., $P_{\text{d}}^{(i)} < 1$, and false alarms are present (cf. Sec.~\ref{sec:sim_results_non_ideal}). For the ideal scenario, we also consider the performance of the stand-alone ARCE localization algorithm at each time
scan.

\subsection{Results in Ideal Scenario}
\label{sec:sim_results_ideal}

We first analyze the ideal scenario, in which the active radar and the receivers do not produce any false alarms and no missed detections are present, that is the detection probability $P_{\text{d}}^{(i)} = P_{\text{d}}$ of each receiver $i$ is equal to one. We consider target 1 and target 2 in two distinct single-target scenarios; both targets start at a range of 30 km. We perform simulations for three different SNR noise levels at $30$ km (i.e., at the beginning of the simulation) equal for all receivers and the active radar, i.e., $\text{SNR}_i = \overline{\text{SNR}}$ for $i=1,\ldots,5$, with $\overline{\text{SNR}} = 0$ dB, $ \overline{\text{SNR}} = -10$, dB and $\overline{\text{SNR}} =-20$ dB.
Figures~\ref{fig:comp-ideal-scenario-border-pos} and~\ref{fig:comp-ideal-scenario-central-pos} show, respectively for target 1 and target 2, the comparison between the ARCE localization algorithm (`ARCE'), the baseline SPA-based MTT algorithm (`SPA-MTT'),
and the proposed algorithm (`Prop.')
both NAD and AD versions with $\tilde{C} = 1,2,3$, in terms of the mean GOSPA (MGOSPA) error, i.e., averaged over the 200 Monte Carlo runs, and for the different values of  $\overline{\text{SNR}}$.
The number of particles $N_\text{p}$ is
500,
$\alpha_\text{r} = 0.7$,
and $\sigma_{\text{ARCE}}$ is set to 500 m.
We first focus on low SNR levels, i.e., $\overline{\text{SNR}} = -10$ dB and   $\overline{\text{SNR}} = -20$ dB. In these cases, one can observe that the ARCE
localization algorithm performs generally worse than the baseline SPA-based MTT technique. This is expected since
ARCE
relies only on single time scan highly noisy measurements without taking advantage of past information. At the same time, the proposed methods are those performing generally better. In particular for target 1 moving close to the antenna's beam edge (Figs.~\ref{fig:comp-ideal-scenario-border-pos_-20db} and \ref{fig:comp-ideal-scenario-border-pos_-10db})
the proposed method leverages the prior information about the
antenna beamwidth of the transmitter by preventing the estimated target to be initialized or to move outside the antenna beam.
The case of target 2, moving in the middle of the antenna's beam, and $\overline{\text{SNR}} = -20$ dB (Fig.~\ref{fig:comp-ideal-scenario-central-pos_-20db}), represents the most challenging scenario and no significant improvement is observed.

Focusing now on the higher  $\overline{\text{SNR}} = 0$ dB, one can observe instead that ARCE exhibits a  better 
performance than the baseline SPA-based MTT method in the first half of the simulations, i.e., from time step $k = 1$ to time step $k = 55$. 
Afterwards, the baseline SPA-based MTT approach becomes more effective. This is probably because of the small number of particles inaccurately representing the prior pdf of the initialized potential target and their propagation over time via the target dynamics. 
This scenario confirms that ARCE can then provide an improved sampling 
of the target space in the initialization phase. 
In fact, the proposed methods are those performing generally better. In particular, one can observe that the NAD approach
provides better results than the AD approach.  This is reasonable since the AD approach constraints the ARCE estimate to lie within a bespoke virtual beam, which might be biased by the target predicted particles and their uncertainty.

In order to comprehend the main advantage of using ARCE to enhance the SPA-based MTT method, we focus on an intermediate SNR level, i.e., $\overline{\text{SNR}} = -10$ dB, and compare the performance of the baseline  SPA-based MTT algorithm and the proposed method --- AD with $\tilde{C}=3$ --- for a varying number of particles in Fig.~\ref{fig:perf_init_final}.
Specifically, this figure shows their
MGOSPA errors averaged over two distinct time intervals, i.e., \textit{interval A} from time
scan 10 to time
scan 40 (continuous lines), and \textit{interval B} from time
scan 41 to time
scan 100 (dashed lines), and on
varying the number of particles $N_\text{p}$.
The performance of the ARCE localization algorithm (`ARCE') is reported for reference, noting that this is independent of the number of particles. 
Top figure shows the results for target 1 (moving close to the antenna's beam edge), while the bottom figure shows the results for target 2 (moving in the middle of antenna's beam).
In both cases, the
largest improvement of the proposed algorithm against the SPA-based MTT algorithm is
achieved for a lower number of particles,
i.e., $N_\text{p} = 500$ or  $N_\text{p} = 1000$, and
for interval A, as shown by the blue and
pink continuous lines.
For interval B and $N_\text{p} = 500$ or  $N_\text{p} = 1000$, the gap in the performance between the SPA-based MTT algorithm and the proposed method is reduced, as shown by the blue and pink dashed lines. 
As the number of particles $N_\text{p}$ increases up to 2500, the SPA-based MTT and the proposed method tend to be equally effective in both intervals A and B.
This behavior suggests that the ARCE estimates provide useful hints for an effective sampling of the space, in particular when targets are initialized (i.e., within time interval A) and for a low number of particles. 
For a larger number of particles, instead, the sampling of the space is inherently more effective, making the impact of the ARCE estimates less significant.
Overall, the use of a lower number of particles is desirable especially when a large number of targets needs to be tracked.

\subsection{Results in Non-Ideal Scenario}
\label{sec:sim_results_non_ideal}

\begin{figure}[!t]
	\centering
	\subfloat[$\overline{\text{SNR}} = -20$ dB.]{
		\includegraphics{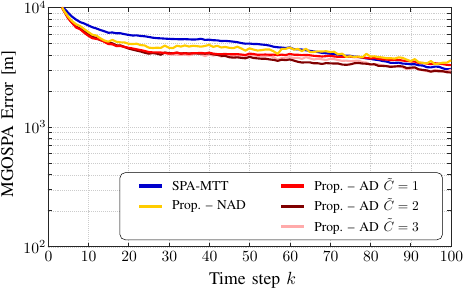}
		\label{fig:comp-not-ideal-scenario-two-targets_-20db}
	}
	
	\centering
	\subfloat[$\overline{\text{SNR}} = -10$ dB.]{
		\includegraphics{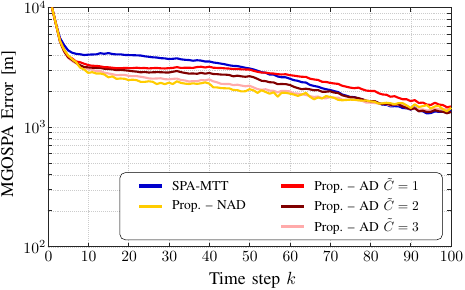}
		\label{fig:comp-not-ideal-scenario-two-targets_-10db}
	}

	\centering
	\subfloat[$\overline{\text{SNR}} = 0$ dB.]{
		\includegraphics{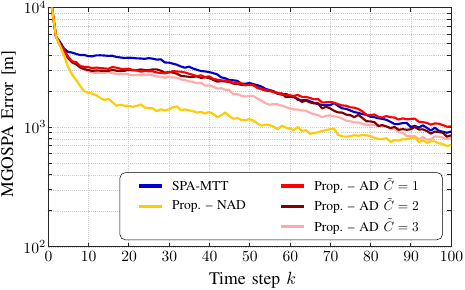}
		\label{fig:comp-not-ideal-scenario-two-targets_0db}
	}
	
	\caption{Comparison between the 
		the baseline SPA-based MTT algorithm (`SPA-MTT') and the proposed algorithm (`Prop.') --- both NAD and AD --- in a non-ideal scenario with two targets moving within the antenna's beam, in terms of MGOSPA error and on varying $\overline{\text{SNR}}$.}
	\label{fig:comp-not-ideal-scenario-two-targets}
\end{figure}

We then analyze a multitarget scenario with both target 1 and target 2, clutter-generated measurements, and 
missed detections. In this scenario, target 1 starts at a range of 35 km, while target 2 at a range of 30 km. The bistatic range of a false alarm  generated by receiver $i$ is linearly distributed between a minimum value equal to the distance $\norm{\V{t}-\V{r}^{(i)}}$ and a maximum value set to 70 km. The number of false alarms for each receiver $i$ is 
modeled according to a Poisson distribution with mean 1, while the detection probability $P_\text{d}^{(i)}$ is 
equal to 0.9 for the active radar $i=1$, and 0.7 for the other receivers $i=2,\ldots,5$. 
The performance of the baseline SPA-based MTT algorithm and the proposed algorithm, both NAD and AD versions with $\tilde{C} = 1,2,3$, is shown in Fig.~\ref{fig:comp-not-ideal-scenario-two-targets}
in terms of MGOSPA error
 versus $\overline{\text{SNR}}$.
As before,
we set $N_\text{p} = 500$,
$\alpha_\text{r} = 0.7$,
and $\sigma_{\text{ARCE}} = 500$ m.
We observe that the proposed NAD version and AD version with $\tilde{C}=3$ still exhibit better
performance than the baseline SPA-based MTT algorithm, especially with
$\overline{\text{SNR}}$ equals to 0 dB and -10 dB.

\section{Conclusions}
\label{sec:conclusions}

Multi-platform radar networks (MPRNs) are becoming an emerging technology due to their capacity of providing improved surveillance capabilities with respect to monostatic and bistatic systems. Due to the rapid
ascent of this technology, there is the need of developing detection, localization, and multi-target tracking (MTT) algorithms to fuse the information obtained by multiple receivers in an efficient way. This article has provided an overview on the most recent localization and tracking techniques for MPRNs. In particular, we have put emphasis on the recently developed angular and range constrained estimator (ARCE) localization algorithm, which exploits the knowledge of the active radar beamwidth, and the scalable sum-product algorithm (SPA) based MTT approach. A solution to combine ARCE with the SPA-based MTT has been introduced in order to exploit the information provided by an active radar beamwidth in 3D MTT scenarios. Finally, experimental results in a simulated 3D scenario have shown that the proposed solution is able to achieve superior performance than the baseline SPA-based MTT.

Possible future
research might regard the extension
of the proposed SPA-based MTT approach combined with ARCE to systems comprising multiple transmitters and in scenarios including multipath environments.

\balance
\bibliographystyle{IEEEtran}
\bibliography{}

\begin{thebibliography}{10}
\providecommand{\url}[1]{#1}
\csname url@samestyle\endcsname
\providecommand{\newblock}{\relax}
\providecommand{\bibinfo}[2]{#2}
\providecommand{\BIBentrySTDinterwordspacing}{\spaceskip=0pt\relax}
\providecommand{\BIBentryALTinterwordstretchfactor}{4}
\providecommand{\BIBentryALTinterwordspacing}{\spaceskip=\fontdimen2\font plus
\BIBentryALTinterwordstretchfactor\fontdimen3\font minus
  \fontdimen4\font\relax}
\providecommand{\BIBforeignlanguage}[2]{{%
\expandafter\ifx\csname l@#1\endcsname\relax
\typeout{** WARNING: IEEEtran.bst: No hyphenation pattern has been}%
\typeout{** loaded for the language `#1'. Using the pattern for}%
\typeout{** the default language instead.}%
\else
\language=\csname l@#1\endcsname
\fi
#2}}
\providecommand{\BIBdecl}{\relax}
\BIBdecl

\bibitem{BraGagSolMenFerLepNicWilBraWin}
M.~Brambilla, D.~Gaglione, G.~Soldi, R.~Mendrzik, G.~Ferri, K.~Lepage,
  M.~Nicoli, P.~Willett, P.~Braca, and M.~Win, ``Cooperative localization and
  multitarget tracking in agent networks with the sum-product algorithm,''
  \emph{IEEE Open Journal of Signal Processing}, vol.~3, pp. 169--195, Mar.
  2022. doi: 10.1109/OJSP.2022.3154684

\bibitem{AUV_Access_2019}
H.~Shakhatreh, A.~H. Sawalmeh, A.~Al-Fuqaha, Z.~Dou, E.~Almaita, I.~Khalil,
  N.~S. Othman, A.~Khreishah, and M.~Guizani, ``Unmanned aerial vehicles
  ({UAVs}): {A} survey on civil applications and key research challenges,''
  \emph{IEEE Access}, vol.~7, pp. 48\,572--48\,634, Apr. 2019. doi:
  10.1109/ACCESS.2019.2909530

\bibitem{hassanalian2017classifications}
M.~Hassanalian and A.~Abdelkefi, ``Classifications, applications, and design
  challenges of drones: {A} review,'' \emph{Progress in Aerospace Sciences},
  vol.~91, pp. 99--131, May 2017.

\bibitem{ferri2017cooperative}
G.~Ferri, A.~Munaf{\`o}, A.~Tesei, P.~Braca, F.~Meyer, K.~Pelekanakis,
  R.~Petroccia, J.~Alves, C.~Strode, and K.~LePage, ``Cooperative robotic
  networks for underwater surveillance: {An} overview,'' \emph{IET Radar Sonar
  Navig.}, vol.~11, no.~12, pp. 1740--1761, Dec. 2017. doi:
  10.1049/iet-rsn.2017.0074

\bibitem{Zeng_16}
Y.~Zeng, R.~Zhang, and T.~J. Lim, ``Wireless communications with unmanned
  aerial vehicles: {Opportunities} and challenges,'' \emph{{IEEE} Commun.
  Mag.}, vol.~54, no.~5, pp. 36--42, May 2016. doi: 10.1109/MCOM.2016.7470933

\bibitem{DIFFUSION2015}
P.~Braca, R.~Goldhahn, G.~Ferri, and K.~LePage, ``Distributed information
  fusion in multistatic sensor networks for underwater surveillance,''
  \emph{{IEEE} Sensors J.}, vol.~16, no.~11, pp. 4003--414, May 2015. doi:
  10.1109/JSEN.2015.2431818

\bibitem{de2021understanding}
J.~Diaz~de Leon, ``Understanding multi-domain operations in {NATO},'' \emph{The
  Three Swords Magazine}, vol.~37, pp. 91--94, 2021.

\bibitem{Pastore_2017}
T.~Pastore, G.~Galdorisi, and A.~Jones, ``Command and control ({C2}) to enable
  multi-domain teaming of unmanned vehicles ({UxVs}),'' in \emph{Proc.
  OCEANS-17}, Anchorage, AK, USA, Sep. 2017.

\bibitem{Che:B98}
V.~S. Chernyak, \emph{Fundamentals of Multisite Radar Systems: Multistatic
  Radars and Multiradar Systems}.\hskip 1em plus 0.5em minus 0.4em\relax
  Amsterdam, The Netherlands: Gordon and Breach, 1998.

\bibitem{OgaDouIng:B18}
D.~W. O’Hagan, S.~R. Doughty, and M.~R. Inggs, ``{Multistatic Radar
  Systems},'' in \emph{Academic Press Library in Signal Processing, Volume 7},
  R.~Chellappa and S.~Theodoridis, Eds.\hskip 1em plus 0.5em minus 0.4em\relax
  Academic Press, 2018, pp. 253--275.

\bibitem{hartley2020cognitive}
D.~S. Hartley~III and K.~O. Jobson, \emph{Cognitive Superiority: Information to
  Power}.\hskip 1em plus 0.5em minus 0.4em\relax Cham, Switzerland: Springer
  Nature, 2020.

\bibitem{AubryCDP19_55}
A.~Aubry, V.~Carotenuto, A.~De~Maio, and L.~Pallotta, ``High range resolution
  profile estimation via a cognitive stepped frequency technique,''
  \emph{{IEEE} Trans. Aerosp. Electron. Syst.}, vol.~55, no.~1, pp. 444--458,
  2019. doi: 10.1109/TAES.2018.2880024

\bibitem{WicMoo}
M.~C. Wicks and W.~Moore, ``Distributed and layered sensing,'' in \emph{Proc.
  International Waveform Diversity and Design Conference}, Pisa, Italy, Jun.
  2007. doi: 10.1109/WDDC.2007.4339417 pp. 233--239.

\bibitem{Lomb}
M.~A. Lombardi, ``The use of {GPS} disciplined oscillators as primary frequency
  standards for calibration and metrology laboratories,'' \emph{NCSLI Measure:
  The Journal of Measurement Science}, vol.~3, no.~3, pp. 56--65, Sep. 2008.
  doi: 10.1080/19315775.2008.11721437

\bibitem{SanIng}
{J. S. Sandenbergh and M. R. Inggs}, ``A common view {GPSDO} to synchronize
  netted radar,'' in \emph{Proc. IET International Conference on Radar
  Systems}, Edinburgh, UK, Oct. 2007.

\bibitem{GriBak}
H.~D. {Griffiths} and C.~J. {Baker}, ``Towards the intelligent adaptive radar
  network,'' in \emph{Proc. IEEE Radar Conference}, Ottawa, Canada, Apr. 2013.
  doi: 10.1109/RADAR.2013.6586005

\bibitem{AubryBDFS23_71}
A.~Aubry, P.~Babu, A.~De~Maio, G.~Fatima, and N.~Sahu, ``A robust framework to
  design optimal sensor locations for {TOA} or {RSS} source localization
  techniques,'' \emph{{IEEE} Trans. Signal Process.}, vol.~71, pp. 1293--1306,
  2023. doi: 10.1109/TSP.2023.3262182

\bibitem{DeMaioBook}
F.~Gini, A.~De~Maio, and L.~Patton, \emph{Waveform Design and Diversity for
  Advanced Radar Systems}, ser. Radar, Sonar and Navigation.\hskip 1em plus
  0.5em minus 0.4em\relax Inst. of Eng. and Technol., 2012.

\bibitem{YiangADYC22_70}
J.~Yang, A.~Aubry, A.~De~Maio, X.~Yu, and G.~Cui, ``Multi-spectrally
  constrained transceiver design against signal-dependent interference,''
  \emph{{IEEE} Trans. Signal Process.}, vol.~70, pp. 1320--1332, 2022. doi:
  10.1109/TSP.2022.3144953

\bibitem{DerDouWooBak:J07}
T.~E. {Derham}, S.~{Doughty}, K.~{Woodbridge}, and C.~J. {Baker}, ``Design and
  evaluation of a low-cost multistatic netted radar system,'' \emph{IET Radar,
  Sonar Navig.}, vol.~1, no.~5, pp. 362--368, Oct. 2007. doi:
  10.1049/iet-rsn:20060100

\bibitem{IngGriFioRitWoo:C14}
M.~Inggs, H.~Griffiths, F.~Fioranelli, M.~Ritchie, and K.~Woodbridge,
  ``Multistatic radar: System requirements and experimental validation,'' in
  \emph{Proc. IEEE International Conference on Radar}, Lille, France, Oct.
  2014.

\bibitem{IngLLewPalRitGri:C19}
M.~R. Inggs, S.~Lewis, R.~Palam\`a, M.~A. Ritchie, and H.~Griffiths, ``Report
  on the 2018 trials of the multistatic {NeXtRAD} dual band polarimetric
  radar,'' in \emph{Proc. IEEE National Conference on Radar}, Boston, MA, USA,
  Apr. 2019.

\bibitem{DouWooBak:C07}
S.~Doughty, K.~Woodbridge, and C.~Baker, ``Improving resolution using
  multistatic radar,'' in \emph{Proc. IET International Conference on Radar
  Systems}, Edinburgh, UK, Oct. 2007.

\bibitem{Griffiths14}
A.-H.~D. Griffiths, ``Keynote address: ``{C}lutter diversity: {A} new degree of
  freedom in multistatic radar'','' in \emph{Proc. {IEEE} Radar Conf.}, 2014.
  doi: 10.1109/RADAR.2014.6875515 pp. 11--11.

\bibitem{NickelGLGK17_2}
B.-R. Klemm, U.~Nickel, C.~Gierull, P.~Lombardo, H.~Griffiths, and W.~Koch,
  \emph{Novel Radar Techniques and Applications: Waveform Diversity and
  Cognitive Radar, and Target Tracking and Data Fusion}, ser. Radar, Sonar and
  Navigation.\hskip 1em plus 0.5em minus 0.4em\relax Inst. of Eng. and
  Technol., 2017, vol.~2.

\bibitem{AubryCDF23_59}
A.~Aubry, V.~Carotenuto, A.~De~Maio, and F.~Fioranelli, ``Compatibility
  assessment of multistatic/polarimetric clutter data with the {SIRP} model,''
  \emph{{IEEE} Trans. Aerosp. Electron. Syst.}, vol.~59, no.~1, pp. 359--374,
  Feb. 2023. doi: 10.1109/TAES.2022.3184916

\bibitem{BeaRitGriMicIngLewKah:C20}
P.~Beasley, M.~Ritchie, H.~Griffiths, W.~Miceli, M.~Inggs, S.~Lewis, and
  B.~Kahn, ``Multistatic radar measurements of {UAVs} at {X}-band and
  {L}-band,'' in \emph{Proc. IEEE Radar Conference}, Florence, Italy, Sep.
  2020.

\bibitem{VivBraGraWil}
G.~Vivone, P.~Braca, K.~Granstrom, and P.~Willett, ``Multistatic {Bayesian}
  extended target tracking,'' \emph{{IEEE} Trans. Aerosp. Electron. Syst.},
  vol.~52, no.~6, pp. 2626--2643, Dec. 2016. doi: 10.1109/TAES.2016.150724

\bibitem{VivBraGraWil2}
G.~Vivone, K.~Granstrom, P.~Braca, and P.~Willett, ``Multiple sensor
  measurement updates for the extended target tracking random matrix model,''
  \emph{{IEEE} Trans. Aerosp. Electron. Syst.}, vol.~53, no.~5, pp. 2544--2558,
  May 2017. doi: 10.1109/TAES.2017.2704166

\bibitem{BracaMAMMW22}
P.~Braca, L.~M. Millefiori, A.~Aubry, S.~Marano, A.~De~Maio, and P.~Willett,
  ``Statistical hypothesis testing based on machine learning: {L}arge
  deviations analysis,'' \emph{IEEE Open J. Signal Process.}, vol.~3, pp.
  464--495, 2022. doi: 10.1109/OJSP.2022.3232284

\bibitem{MeyBraWilHla:J17}
F.~Meyer, P.~Braca, P.~Willett, and F.~Hlawatsch, ``A scalable algorithm for
  tracking an unknown number of targets using multiple sensors,'' \emph{{IEEE}
  Trans. Signal Process.}, vol.~65, no.~13, pp. 3478--3493, Jul. 2017. doi:
  10.1109/TSP.2017.2688966

\bibitem{MeyKroWilLauHlaBraWin:J18}
F.~Meyer, T.~Kropfreiter, J.~L. Williams, R.~A. Lau, F.~Hlawatsch, P.~Braca,
  and M.~Z. Win, ``Message passing algorithms for scalable multitarget
  tracking,'' \emph{Proc. {IEEE}}, vol. 106, no.~2, pp. 221--259, Feb. 2018.
  doi: 10.1109/JPROC.2018.2789427

\bibitem{AubBraDemMar:J22}
A.~Aubry, P.~Braca, A.~{De Maio}, and A.~Marino, ``Enhanced target localization
  with deployable multiplatform radar nodes based on non-convex constrained
  least squares optimization,'' \emph{{IEEE} Trans. Signal Process.}, vol.~70,
  pp. 1282--1294, Feb. 2022. doi: 10.1109/TSP.2022.3147037

\bibitem{WilGri:B07}
N.~J. Willis and H.~D. Griffiths, \emph{Advances in Bistatic Radar}.\hskip 1em
  plus 0.5em minus 0.4em\relax Raleigh, NC, USA: SciTech Publishing, 2007.

\bibitem{MalKul:J12}
M.~{Malanowski} and K.~{Kulpa}, ``Two methods for target localization in
  multistatic passive radar,'' \emph{{IEEE} Trans. Aerosp. Electron. Syst.},
  vol.~48, no.~1, pp. 572--580, Jan. 2012. doi: 10.1109/TAES.2012.6129656

\bibitem{GiaCecScoGar}
A.~Giannitrapani, N.~Ceccarelli, F.~Scortecci, and A.~Garulli, ``Comparison of
  {EKF} and {UKF} for spacecraft localization via angle measurements,''
  \emph{{IEEE} Trans. Aerosp. Electron. Syst.}, vol.~47, no.~1, pp. 75--84,
  Jan. 2011. doi: 10.1109/TAES.2011.5705660

\bibitem{UllSheSuEspCho}
I.~Ullah, Y.~Shen, X.~Su, C.~Esposito, and C.~Choi, ``A localization based on
  unscented {Kalman} filter and particle filter localization algorithms,''
  \emph{IEEE Access}, vol.~8, pp. 2233--2246, Dec. 2020. doi:
  10.1109/ACCESS.2019.2961740

\bibitem{BisFidAndDogPat}
A.~N. Bishop, B.~Fidan, B.~D.~O. Anderson, K.~Dogancay, and P.~N. Pathirana,
  ``Optimality analysis of sensor-target localization geometries,''
  \emph{Automatica}, vol.~46, no.~3, p. 479–492, Mar. 2010.

\bibitem{OguGomXvStoOli}
P.~Oğuz-Ekim, J.~P. Gomes, J.~Xavier, M.~Stošić, and P.~Oliveira, ``An
  angular approach for range-based approximate maximum likelihood source
  localization through convex relaxation,'' \emph{{IEEE} Trans. Wireless
  Commun.}, vol.~13, no.~7, pp. 3951--3964, Jul. 2014. doi:
  10.1109/TWC.2014.2314653

\bibitem{DiaTabDiaSed}
M.~Dianat, M.~R. Taban, J.~Dianat, and V.~Sedighi, ``Target localization using
  least squares estimation for {MIMO} radars with widely separated antennas,''
  \emph{{IEEE} Trans. Aerosp. Electron. Syst.}, vol.~49, no.~4, pp. 2730--2741,
  Oct. 2013. doi: 10.1109/TAES.2013.6621849

\bibitem{AmiBehZam}
F.~B. R.~Amiri and H.~Zamani, ``Asymptotically efficient target localization
  from bistatic range measurements in distributed {MIMO} radars,'' \emph{{IEEE}
  Signal Process. Lett.}, vol.~24, no.~3, pp. 299--303, Mar. 2017. doi:
  10.1109/LSP.2017.2660545

\bibitem{AubCarDemPal:J20}
A.~Aubry, V.~Carotenuto, A.~{De Maio}, and L.~Pallotta, ``Localization in {2D
  PBR} with multiple transmitters of opportunity: A constrained least squares
  approach,'' \emph{{IEEE} Trans. Signal Process.}, vol.~68, pp. 634--646, Jan.
  2020. doi: 10.1109/TSP.2020.2964235

\bibitem{AubBraDemMar:J21}
A.~{Aubry}, P.~{Braca}, A.~{De Maio}, and A.~{Marino}, ``{2D PBR} localization
  complying with constraints forced by active radar measurements,''
  \emph{{IEEE} Trans. Aerosp. Electron. Syst.}, vol.~57, no.~5, pp. 2647--2660,
  Oct. 2021. doi: 10.1109/TAES.2021.3067612

\bibitem{Ber:B16}
D.~P. Bertsekas, \emph{Nonlinear Programming}.\hskip 1em plus 0.5em minus
  0.4em\relax Nashua, NH, USA: Athena Scientific, 2016.

\bibitem{SheMolSal:J12}
J.~{Shen}, A.~{Molisch}, and J.~{Salmi}, ``Accurate passive location estimation
  using {TOA} measurements,'' \emph{{IEEE} Trans. Wireless Commun.}, vol.~11,
  no.~6, pp. 2182--2192, Jun. 2012. doi: 10.1109/TWC.2012.040412.110697

\bibitem{ZhaHo:J19}
Y.~{Zhang} and K.~C. {Ho}, ``Multistatic localization in the absence of
  transmitter position,'' \emph{{IEEE} Trans. Signal Process.}, vol.~67,
  no.~18, pp. 4745--4760, Sep. 2019. doi: 10.1109/TSP.2019.2929960

\bibitem{BarWilTia:B11}
Y.~Bar-Shalom, P.~K. Willett, and X.~Tian, \emph{{Tracking and Data Fusion: A
  Handbook of Algorithms}}.\hskip 1em plus 0.5em minus 0.4em\relax Storrs, CT,
  USA: YBS Publishing, 2011.

\bibitem{DezBar:TR21}
J.~Dezert and Y.~Bar-Shalom, ``Computational complexity of {JPDA}: Worst case
  analysis,'' Office National d'Etudes et de Recherches A\'erospatiales
  (ONERA), Tech. Rep. TN-2021-04-21, Apr. 2021.

\bibitem{Mah:B07}
R.~Mahler, \emph{{Statistical Multisource-Multitarget Information
  Fusion}}.\hskip 1em plus 0.5em minus 0.4em\relax Norwood, MA, USA: Artech
  House, 2007.

\bibitem{ChaMor:B11}
S.~Challa, M.~R. Morelande, D.~Mu{\v s}icki, and R.~J. Evans,
  \emph{{Fundamentals of Object Tracking}}.\hskip 1em plus 0.5em minus
  0.4em\relax Cambridge, UK: Cambridge University Press, 2011.

\bibitem{MusEva:J04}
D.~{Mu{\v s}icki} and R.~{Evans}, ``Joint integrated probabilistic data
  association: {JIPDA},'' \emph{{IEEE} Trans. Aerosp. Electron. Syst.},
  vol.~40, no.~3, pp. 1093--1099, Sep. 2004. doi: 10.1109/TAES.2004.1337482

\bibitem{Rei:J79}
D.~B. Reid, ``An algorithm for tracking multiple targets,'' \emph{{IEEE} Trans.
  Autom. Control}, vol.~24, no.~6, pp. 843--854, Dec. 1979. doi:
  10.1109/TAC.1979.1102177

\bibitem{ChoMorRei:J19}
C.-Y. Chong, S.~Mori, and D.~B. Reid, ``Forty years of multiple hypothesis
  tracking,'' \emph{J. Adv. Inf. Fusion}, vol.~14, no.~2, pp. 131--153, Dec.
  2019. doi: 10.23919/ICIF.2018.8455386

\bibitem{Mah:J03}
R.~P.~S. {Mahler}, ``Multitarget {Bayes} filtering via first-order multitarget
  moments,'' \emph{{IEEE} Trans. Aerosp. Electron. Syst.}, vol.~39, no.~4, pp.
  1152--1178, Oct. 2003. doi: 10.1109/TAES.2003.1261119

\bibitem{Mah:J07}
R.~{Mahler}, ``{PHD} filters of higher order in target number,'' \emph{{IEEE}
  Trans. Aerosp. Electron. Syst.}, vol.~43, no.~4, pp. 1523--1543, Oct. 2007.
  doi: 10.1109/TAES.2007.4441756

\bibitem{VoVoCan:J07}
B.-T. Vo, B.-N. Vo, and A.~Cantoni, ``Analytic implementations of the
  cardinalized probability hypothesis density filter,'' \emph{{IEEE} Trans.
  Signal Process.}, vol.~55, no.~7, pp. 3553--3567, Jul. 2007. doi:
  10.1109/TSP.2007.894241

\bibitem{NagCla:C11}
S.~{Nagappa} and D.~E. Clark, ``On the ordering of the sensors in the
  iterated-corrector probability hypothesis density {(PHD)} filter,'' in
  \emph{Proc. SPIE-11}, vol. 8050, Orlando, FL, USA, Apr. 2011, pp. 26--28.

\bibitem{NanBloCoaRab:J16}
S.~Nannuru, S.~Blouin, M.~Coates, and M.~Rabbat, ``Multisensor {CPHD} filter,''
  \emph{{IEEE} Trans. Aerosp. Electron. Syst.}, vol.~52, no.~4, pp. 1834--1854,
  Aug. 2016. doi: 10.1109/TAES.2016.150265

\bibitem{BraMarMatWil:J13}
P.~{Braca}, S.~{Marano}, V.~{Matta}, and P.~{Willett}, ``Asymptotic efficiency
  of the {PHD} in multitarget/multisensor estimation,'' \emph{{IEEE} J. Sel.
  Topics Signal Process.}, vol.~7, no.~3, pp. 553--564, Apr. 2013. doi:
  10.1109/JSTSP.2013.2257161

\bibitem{VoVo:J13}
B.-T. Vo and B.-N. Vo, ``Labeled random finite sets and multi-object conjugate
  priors,'' \emph{{IEEE} Trans. Signal Process.}, vol.~61, no.~13, pp.
  3460--3475, Jul. 2013. doi: 10.1109/TSP.2013.2259822

\bibitem{ReuVoVoDie:J14}
S.~{Reuter}, B.~{Vo}, B.~{Vo}, and K.~{Dietmayer}, ``The labeled
  multi-{B}ernoulli filter,'' \emph{{IEEE} Trans. Signal Process.}, vol.~62,
  no.~12, pp. 3246--3260, May 2014. doi: 10.1109/TSP.2014.2323064

\bibitem{SolMeyBraHla:J19}
G.~Soldi, F.~Meyer, P.~Braca, and F.~Hlawatsch, ``Self-tuning algorithms for
  multisensor-multitarget tracking using belief propagation,'' \emph{{IEEE}
  Trans. Signal Process.}, vol.~67, no.~15, pp. 3922--3937, Aug. 2019. doi:
  10.1109/TSP.2019.2916764

\bibitem{GagSolMeyHlaBraFarWin:J20}
D.~Gaglione, G.~Soldi, F.~Meyer, F.~Hlawatsch, P.~Braca, A.~Farina, and M.~Z.
  Win, ``Bayesian information fusion and multitarget tracking for maritime
  situational awareness,'' \emph{IET Radar, Sonar Navig.}, vol.~14, no.~12, p.
  1845–1857, Dec. 2020. doi: 10.1049/iet-rsn.2019.0508

\bibitem{GagBraSolMeyHlaMoe:J22}
D.~Gaglione, P.~Braca, G.~Soldi, F.~Meyer, F.~Hlawatsch, and M.~Z. Win,
  ``Fusion of sensor measurements and target-provided information in
  multitarget tracking,'' \emph{{IEEE} Trans. Signal Process.}, vol.~70, pp.
  322--336, Dec. 2021. doi: 10.1109/TSP.2021.3132232

\bibitem{Poo:B94}
H.~V. Poor, \emph{An Introduction to Signal Detection and Estimation},
  2nd~ed.\hskip 1em plus 0.5em minus 0.4em\relax New York, NY, USA: Springer,
  1994.

\bibitem{LiBolDju:J15}
T.~Li, M.~Bolic, and P.~M. Djuric, ``Resampling methods for particle filtering:
  Classification, implementation, and strategies,'' \emph{{IEEE} Signal
  Process. Mag.}, vol.~32, no.~3, pp. 70--86, Apr. 2015. doi:
  10.1109/MSP.2014.2330626

\bibitem{AruMasGorCla:J02}
M.~S. Arulampalam, S.~Maskell, N.~Gordon, and T.~Clapp, ``A tutorial on
  particle filters for online nonlinear/non-{G}aussian {B}ayesian tracking,''
  \emph{{IEEE} Trans. Signal Process.}, vol.~50, no.~2, pp. 174--188, Feb.
  2002. doi: 10.1109/78.978374

\bibitem{CappeGM07_95}
O.~Cappe, S.~J. Godsill, and E.~Moulines, ``An overview of existing methods and
  recent advances in sequential monte carlo,'' \emph{Proc. {IEEE}}, vol.~95,
  no.~5, pp. 899--924, 2007. doi: 10.1109/JPROC.2007.893250

\bibitem{ShaoHL:J10}
X.~Shao, B.~Huang, and J.~M. Lee, ``Constrained {B}ayesian state estimation –
  {A} comparative study and a new particle filter based approach,''
  \emph{Journal of Process Control}, vol.~20, no.~2, pp. 143--157, Feb. 2010.
  doi: 10.1016/j.jprocont.2009.11.002

\bibitem{RahGarSven:C17}
A.~S. {Rahmathullah}, {\'A}.~F. {Garc{\'i}a-Fern{\'a}ndez}, and L.~{Svensson},
  ``Generalized optimal sub-pattern assignment metric,'' in \emph{Proc.
  FUSION-17}, Xi'an, China, Jul. 2017.

\end{thebibliography}

\end{document}